\documentclass[10pt,conference]{IEEEtran}

\IEEEoverridecommandlockouts
\usepackage{cite}
\usepackage{amsmath,amssymb,amsfonts}
\usepackage{algorithmic}
\usepackage{graphicx}
\usepackage{textcomp}
\usepackage{xcolor}
\usepackage{hyperref}
\usepackage{times}
\usepackage{enumitem}
\usepackage{lipsum}
\usepackage{array}
\def\BibTeX{{\rm B\kern-.05em{\sc i\kern-.025em b}\kern-.08em
    T\kern-.1667em\lower.7ex\hbox{E}\kern-.125emX}}
\usepackage{booktabs} 
\renewcommand{\paragraph}[1]{\vspace{0.3\baselineskip}\noindent\textbf{#1}}
\usepackage{pgfplots, tikz}
\usepackage{pgfplotstable}
\usepackage{array}
\usepackage{booktabs}
\usepackage{comment}
\usepackage{todonotes}
\usetikzlibrary{pgfplots.groupplots}
\usetikzlibrary{shapes.geometric}
\usetikzlibrary{positioning,fit,shapes.geometric,backgrounds, calc}
\usetikzlibrary{arrows,decorations.markings}
\usetikzlibrary{shapes.arrows}
\usetikzlibrary{patterns}
\tikzset{textnode/.style={inner sep=0pt,outer sep=0,execute at begin node={\strut}}}
\tikzstyle{state} = [textnode,circle, draw, inner sep=0pt, outer sep=0]

\usepackage{helvet}

\pgfplotsset{every axis/.append style={
  tick label style = {font=\scriptsize\sffamily},
  legend style = {font=\footnotesize\sffamily},
  title style = {font=\footnotesize\sffamily},
  label style = {font=\footnotesize\sffamily}
                    }}

\selectcolormodel{cmyk}
\usepgfplotslibrary{colorbrewer}

\usepackage{mathtools}

\usepackage{multirow}
\usepackage{graphicx}
\usepackage{listings}

\newenvironment{customlegend}[1][]{%
    \begingroup
    \csname pgfplots@init@cleared@structures\endcsname
    \pgfplotsset{#1}%
}{%
    \csname pgfplots@createlegend\endcsname
    \endgroup
}%

\def\addlegendimage{\csname pgfplots@addlegendimage\endcsname}

\pgfplotsset{compat=1.14,
    /pgfplots/ybar legend/.style={
    /pgfplots/legend image code/.code={%
       \draw[##1,/tikz/.cd,yshift=-0.25em]
        (0cm,0cm) rectangle (3pt,0.8em);},
   },
}

\newcommand{\nop}[1]{}

\newcommand*{\eg}{{\em e.g.}}

\newcommand*{\ie}{{\em i.e.}}

\newcommand{\rk}[1]{\textcolor{blue}{RK: #1}}

\usepackage[T1]{fontenc}
\usepackage{etoolbox}

\begin{document}
\title{\fontsize{23}{30}\selectfont Modelling Online Comment Threads from their Start}


\author{
\IEEEauthorblockN{Rachel Krohn \qquad Tim Weninger}
\IEEEauthorblockA{\textit{Department of Computer Science and Engineering} \\
\textit{University of Notre Dame}\\
Notre Dame, IN, USA \\
\texttt{\{rkrohn,tweninge\}@nd.edu}} 
}

\maketitle

\begin{abstract}
The social Web is a widely used platform for online discussion. Across social media, users can start discussions by posting a topical image, url, or message. Upon seeing this initial post, other users may add their own comments to the post, or to another user's comment. The resulting online discourse produces a \textit{comment thread}, which constitutes an enormous portion of modern online communication. Comment threads are often viewed as trees: nodes represent the post and its comments, while directed edges represent reply-to relationships. The goal of the present work is to predict the size and shape of these comment threads. Existing models do this by observing the first several comments and then fitting a predictive model. However, most comment threads are relatively small, and waiting for data to materialize runs counter to the goal of the prediction task. We therefore introduce the Comment Thread Prediction Model (CTPM) that accurately predicts the size and shape of a comment thread using only the text of the initial post, allowing for the prediction of new posts without observable comments. We find that the CTPM significantly outperforms existing models and competitive baselines on thousands of Reddit discussions from nine varied subreddits, particularly for new posts.
\end{abstract}

\maketitle

\section{Introduction}

Online forums are an important source for discussion and information sharing. Commentary on social media posts color the presentation of the content and allow the content consumer to ask questions, engage in debate, or otherwise contribute to discussion on the topic. The nature of online social commentary has been widely studied across various contexts. Research in cyber-bullying and hate speech seeks to understand and reduce anti-social behavior, political and social scientists seek to glean public opinion from social commentary, and data scientists seek to use the structure of discussions as a way to organize information. The vast amount of publicly available, born-digital, human-generated discussion stands to significantly contribute to our understanding of information, knowledge, and social order.

Online discussions take many forms. Some communication, like emails and texts, allow a group of individuals to privately share information in a linear fashion. Social networks, however, often exhibit non-linear, public discussions between thousands of users simultaneously. Cascades of Twitter and Facebook posts have received significant attention due to their enormous popularity~\cite{gao2015modeling}. Typical research in these topics seeks to simulate the spread of a news article, video, or meme, and has recently been re-invigorated due to a renewed interest in the spread of misinformation. Understanding resharing and retweeting behavior is an important topic, but little attention has been given to predicting the shape of the discussion threads that accompany these social posts.

This is an extremely difficult task - it may not be possible to make accurate predictions at all~\cite{salganik2006experimental,glenski2018guessthekarma,glenski2017rating}. The distribution of comment thread sizes has a very long tail, further complicating the prediction task \cite{goel2012structure}.

Each online social system is slightly different, but generally discussion threads are attached to a post. Each post has a headline or title, a timestamp, and a poster (\ie, the user who submitted the post to the system), along with an associated discussion thread. The thread itself may have one or more top-level comments that express some topically relevant fact or opinion. Each top-level comment may motivate further discussion in the form of children, grandchildren, etc. comments. In this way, a discussion thread resembles a tree, where the post itself is the root and the comments and replies represent the various branches and leaves of the tree.

\vspace{.2cm}
\noindent\textbf{Present Work.} The primary goal of the present work is to predict the shape and size of these discussions. We also endeavour to accurately predict all sizes of comment threads, and to make predictions for new posts without comments.

We focus primarily on Reddit, but the dynamics that create discussion threads appear in many domains. However, the driving mechanisms can rely heavily on the underlying social network, like on Facebook and Twitter. Online discussion boards like Reddit, Digg, and YouTube are significantly different because no social network exists. Instead, any user may access and contribute to any discussion\cite{choi2015characterizing}. 

Popularity prediction models have been established using the Hawkes process~\cite{medvedev2018modelling}, linear regression~\cite{szabo2008predicting}, deep reinforcement learning~\cite{he2016deep}, or entity linking and LSTMs~\cite{dou2018predicting}. Generally speaking, the methodology of these studies watches the first several hours of a thread (or Twitter cascade) and then constructs a model to predict the remainder of the thread. The first several hours are critical because early attention is strongly correlated to high popularity~\cite{glenski2017rating}. Instead, we ask a more difficult question: Can final cascade size and shape be predicted when none or only a few comments are available? 

To answer this question, we introduce the Comment Thread Prediction Model (CTPM) that uses the mechanics of graph representation learning algorithms like node2vec~\cite{grover2016node2vec} to fit the parameters of a Hawkes process. 

\vspace{.2cm}
\noindent\textbf{Main Findings.} 
Exhaustive experiments over millions of comments show CTPM is able to 1) quickly learn complex, human-interpretable parameters, and 2) accurately predict the size and shape of online discussion threads from only the post title and the submitting user, before comments are present.

\section{Preliminaries}

Before we introduce the CTPM, this section presents several foundational concepts. 

Point processes are a sequence of discrete, inter-dependent, points within a continuous space. Temporal point processes, therefore, are discrete events that occur in a continuous time. Stochastic events, like earthquakes and stock market volatility, can be modelled by point processes, which are typically defined as follows.

Let $T_1$, $T_2$, $\ldots$ be discrete event times, where the $i^\textrm{th}$ event is typically represented as $T_i$. Let $N_t$ be the count of events that occurred up to time $t$. 

Let $\tau$ be a sequence of exponential random variables defined by parameter $\lambda$; then, $T_n=\sum_1^n{\tau_i}$. Typically, the values of $\tau_1$, $\tau_2$, $\ldots$ are called the {\em inter-arrival times} for a sequence of random events, and are governed by the {\em rate} represented by $\lambda$. Because $\tau$ is a sequence of exponential random variables, we call this a {\em Poisson point process} with a probability density function of $\tau$ w.r.t $t$ of $\lambda e^{-\lambda t}$. 

Simply put, the rate parameter $\lambda$ indicates that events arrive at an average rate of $\lambda$ per unit time. However, $\lambda$ need not be a constant value. Instead, we may wish to define $\lambda^{(t)}$ where the rate parameter changes with $t$. These cases are called {\em non-homogeneous} Poisson point processes, and they are commonly found when one event increases the likelihood that another will occur, as in the case of aftershocks following earthquakes. One of the most well-known of these {\em self-exciting} processes is called the Hawkes process, which takes the following form:

$$\lambda^{(t)} = h(t) + n_b\sum_{i:t>\tau_i}\phi\left(t-\tau_i\right)$$

\noindent where $h$ is the base intensity, $\phi$ is a memory kernel, and $t>\tau_i$ is all of those events that occurred before $t$. This process is called self-exciting because the probability of a new event at time $t$ increases after an event at time $\tau_i$ by the kernel $\phi(t-\tau_i)$. 

By analogy, events in a Hawkes process consist of two types: `immigrants' which are those events that occur without a preceding/parent event, and `offspring' which are those events that are produced by some previous event. These dynamics create many `family-trees' where immigrants are generated with a rate of $h$, and the branching factor $n_b$ regulates the number of offspring per parent event. Each offspring $T_i$ is generated after $\tau_i$ and can themselves generate offspring with rate $n_b\phi(t-\tau_i)$. 

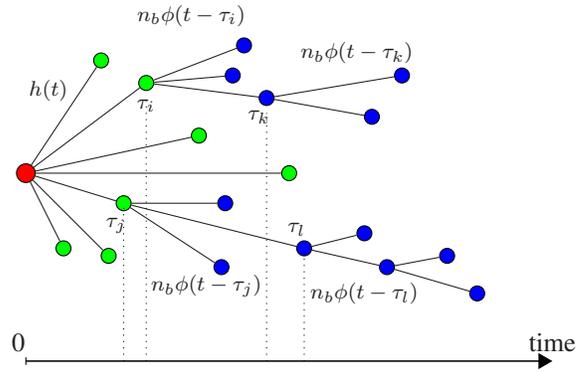
\begin{figure}[t]
    \centering
    \begin{tikzpicture}


\node[shape=circle, inner sep=2.5, draw=black, fill=red] (root) at (0, 0) {};

\node[shape=circle, inner sep=2, draw=black, fill=green] (a) at (0.5,-1) {};
	\path [-] (root) edge (a);
\node[shape=circle, inner sep=2, draw=black, fill=green] (b) at (1.3,-0.4) {};
	\path [-] (root) edge (b);
\node[shape=circle, inner sep=2, draw=black, fill=green] (c) at (1.1,-1.1) {};
	\path [-] (root) edge (c);
\node[shape=circle, inner sep=2, draw=black, fill=green] (d) at (3.5,0) {};
	\path [-] (root) edge (d);
\node[shape=circle, inner sep=2, draw=black, fill=green] (e) at (2.3,0.5) {};
	\path [-] (root) edge (e);
\node[shape=circle, inner sep=2, draw=black, fill=green] (f) at (1.6,1.2) {};
	\path [-] (root) edge (f);
\node[shape=circle, inner sep=2, draw=black, fill=green] (g) at (1, 1.5) {};
	\path [-] (root) edge (g);
	
\node[shape=circle, inner sep=2, draw=black, fill=blue] (1) at (2.6,-1.25) {};
	\path [-] (b) edge (1);
\node[shape=circle, inner sep=2, draw=black, fill=blue] (2) at (3.7,-1) {};
	\path [-] (b) edge (2);
	\node[shape=circle, inner sep=2, draw=black, fill=blue] (2a) at (4.5,-0.8) {};
		\path [-] (2) edge (2a);
	\node[shape=circle, inner sep=2, draw=black, fill=blue] (2b) at (4.8,-1.25) {};
		\path [-] (2) edge (2b);
		\node[shape=circle, inner sep=2, draw=black, fill=blue] (2bi) at (6,-1.6) {};
			\path [-] (2b) edge (2bi);
		\node[shape=circle, inner sep=2, draw=black, fill=blue] (2bii) at (5.6,-1.1) {};
			\path [-] (2b) edge (2bii);
\node[shape=circle, inner sep=2, draw=black, fill=blue] (3) at (2.65,-0.4) {};
	\path [-] (b) edge (3);

\node[shape=circle, inner sep=2, draw=black, fill=blue] (4) at (3.2,1) {};
	\path [-] (f) edge (4);
	\node[shape=circle, inner sep=2, draw=black, fill=blue] (4a) at (4.6,0.75) {};
		\path [-] (4) edge (4a);
	\node[shape=circle, inner sep=2, draw=black, fill=blue] (4a) at (5,1.3) {};
		\path [-] (4) edge (4a);
\node[shape=circle, inner sep=2, draw=black, fill=blue] (5) at (2.75,1.3) {};
	\path [-] (f) edge (5);
\node[shape=circle, inner sep=2, draw=black, fill=blue] (7) at (2.9,1.7) {};
	\path [-] (f) edge (7);
	
\draw[draw=black,-triangle 60,fill=black]  (0,-2.5) -- (7,-2.5);
\draw[draw=black] (0,-2.55) -- (0,-2.45);
\node (axis0) at (-0.1,-2.25) {0};
\node (axistime) at (7,-2.25) {time};

\node (mu) at (.3,1.1) {\footnotesize $h(t)$};

\node(ti) at (1.6,0.9) {\footnotesize $\tau_i$};
\node (phii) at (2.2,2.1) {\footnotesize $n_b\phi(t-\tau_i)$};

\node(tk) at (3.1,0.75) {\footnotesize $\tau_k$};
\node (phik) at (4.4,1.6) {\footnotesize $n_b\phi(t-\tau_k)$};

\node(tj) at (1.2,-0.7) {\footnotesize $\tau_j$};
\node (phij) at (2.4,-1.5) {\footnotesize $n_b\phi(t-\tau_j)$};

\node(tl) at (3.6,-0.75) {\footnotesize $\tau_l$};
\node (phil) at (4.5,-1.6) {\footnotesize $n_b\phi(t-\tau_l)$};

\draw[draw=black, dotted] (1.3, -2.5) -- (1.3, -0.4);
\draw[draw=black, dotted] (1.6, -2.5) -- (1.6, 1.0);
\draw[draw=black, dotted] (3.2, -2.5) -- (3.2, 0.8);
\draw[draw=black, dotted] (3.7, -2.5) -- (3.7, -1);

\draw[draw=white] (0, -2.75) -- (7, -2.75);

\end{tikzpicture}
    \vspace{-3mm}
    \caption{Example of a Hawkes branching process. The red node (far left) represents a social media post. Green and blue nodes represent `immigrant' and 'offspring' events respectively. Each event is generated at $\tau_i$. Immigrant events are sampled from $h(t)$, and offspring events are sampled with intensity $\phi(t-\tau_i)$ and have a branching factor of $n_b$. This Hawkes process model is a natural representation of a discussion cascade. Adapted with permission from Medvedev et al.~\cite{medvedev2018modelling}}
    \label{fig:hawkes_tree}
\end{figure}

Applying this analogy to comments in online discussion forums, Medvedev et al. defined a root node as a social media post, and `immigrants' as top-level comments, which can each have reply-comment `offspring'~\cite{medvedev2018modelling}. The example in Fig.~\ref{fig:hawkes_tree} illustrates a small discussion cascade. The Hawkes process is particularly well-suited to the modelling of discussion threads, and has been previously used to predict the final retweet cascade size and shape~\cite{rizoiu2017tutorial,zhou2013learning} based on user history~\cite{zhao2015seismic}, circadian rhythms~\cite{kobayashi2016tideh}, or other user properties~\cite{yang2010predicting}.

What remains is to define the functional forms of $h$ and $\phi$. Previously, Medvedev et al. demonstrated, based on an examination of common Reddit response times, that $h$ is best articulated as a Weibull distribution \textsf{Weib}$(a,b,\alpha)$ and $\phi$ is represented by the lognormal distribution \textsf{logN}$(\mu,\sigma)$~\cite{medvedev2018modelling}. These distributions hold regardless of the level or age of the comments. With these forms defined, the model parameters $a, b,\alpha,\mu,\sigma$ and the branching factor $n_b$ can be estimated using well defined log-likelihood functions~\cite{daley2007introduction} and an optimization algorithm like BFGS.

Unfortunately, this parameter estimation assumes the existence of data in the form of an existing comment thread: a minimum of 10 observed comments are required to fit parameters, and both root-level and comment replies must be present. Therefore, these models typically estimate the parameters from the first hours of a discussion thread to predict the remainder. Posts without comments cannot be predicted.

Our goal is to estimate $a,b,\alpha,\mu,\sigma, \textrm{and } n_b$ without an existing comment thread. Hence, our inference algorithm provides the ability to simulate discussions, and predict their popularity, from their very beginning.

\section{Comment Thread Prediction Model}

Predicting the shape and size of a discussion thread or any event cascade is extremely challenging, but this task provides a rich avenue for compelling questions:

\begin{enumerate}
    \item Can discussion threads be predicted using only post features?
    \item Can we infer model parameters of a discussion thread that does not yet have discussion?
    \item How much do the early dynamics of a discussion help improve the model?
\end{enumerate}

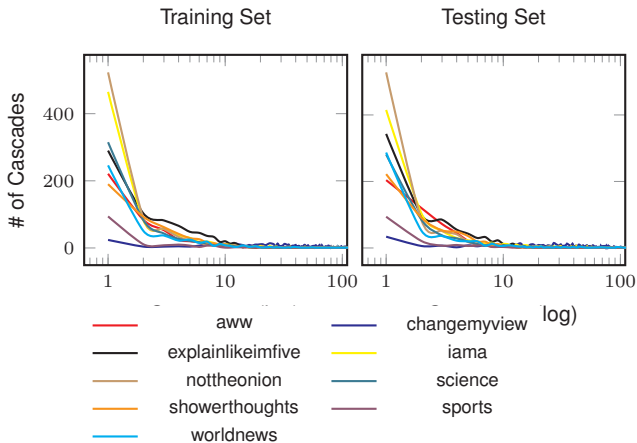
\begin{figure}
    \centering
    \include{./figs/size_dist}
    \vspace{-0.8cm}
    \begin{tikzpicture}
    \begin{customlegend}[ 
    legend columns=2,
  legend style={
    draw=none,
    column sep=2ex,
    font=\sffamily\scriptsize,
  },
  legend entries={\textsf{aww}, \textsf{changemyview}, \textsf{explainlikeimfive}, \textsf{iama}, \textsf{nottheonion}, \textsf{science}, \textsf{showerthoughts}, \textsf{sports}, \textsf{worldnews}},
  ]
    \addlegendimage{red, thick}
    \addlegendimage{blue, thick}
    \addlegendimage{black, thick}
    \addlegendimage{yellow, thick}
    \addlegendimage{brown, thick}
    \addlegendimage{teal, thick}
    \addlegendimage{orange, thick}
    \addlegendimage{violet, thick}
    \addlegendimage{cyan, thick}
    \end{customlegend}
\end{tikzpicture}
    \vspace{-4mm}
    \caption{Cascade size distribution for training and testing data from Reddit. Existing models require at least 10 comments before they can generate predictions; however, we observe that the vast majority of cascades are less than 10 comments.}
    \label{fig:size_dist}
\end{figure}

Typically, models simulate cascades by fitting parameters to early data, but most posts generate little discussion. Data from Reddit discussion threads, shown in Fig.~\ref{fig:size_dist} and described in more detail below, illustrates that most cascades fizzle out after a few comments, although the shape of this distribution depends on the community. This further stresses the need for our cascade prediction model if we wish to predict the vast majority of cascades.

The abundance of small discussions makes existing models impractical for application to a wider set of posts. Instead of depending on observed discussion comments, we rely on parameters fit to previous similar discussions and infer new parameters based on this history. 

We show that this kind of parameter-transfer works reasonably well, though it is important to recognize some of the assumptions that we make in this approach. First, we implicitly assume that the discussion contributed to a post is correlated with features of the post like the title, submitting-user, and community (\ie, subreddit). This can be a difficult assumption to make because identical content is known to receive drastically different attention~\cite{lakkaraju2013s}. Second, we assume that similar posts should receive similar discussion. This can also be a difficult assumption because of the apparent disconnect between social media popularity and user-preference~\cite{glenski2018guessthekarma}.

In the remainder of this section we describe the Comment Thread Prediction Model (CTPM). Our primary contribution is in the parameter inference for posts with few or no observed comments. For this difficult task the CTPM simulates comment cascades attached to new Reddit posts, before comments are present. Source code, data, and evaluation scripts are available on Github\footnote{\url{https://github.com/rkrohn/redditmodel}}. The CTPM contains three primary steps:

\nop{
An overview of CTPM is shown in Fig.~\ref{fig:system_diagram}, which contains three steps:
\begin{figure}[t]
    \centering
    \includegraphics[width=\linewidth]{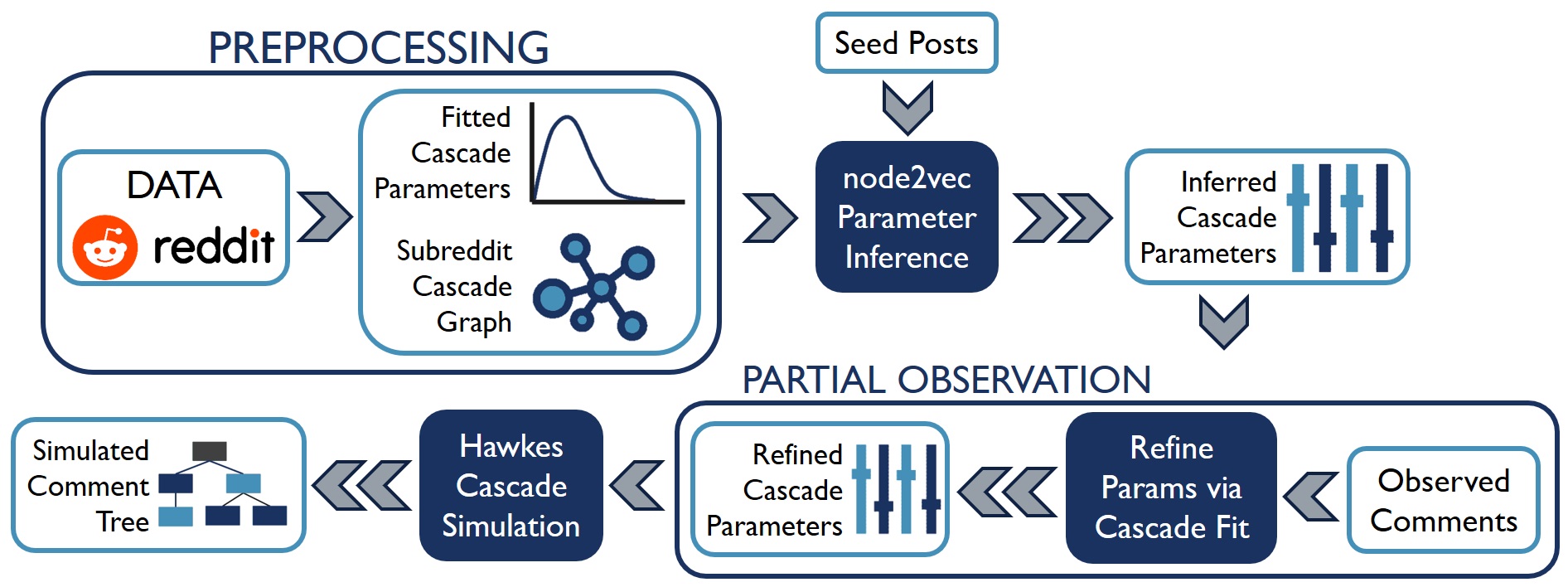}
    \caption{Overall system diagram.}
    \label{fig:system_diagram}
\end{figure}
}

\begin{enumerate}[label=Step~\arabic*:,leftmargin=*,align=left]
\item Preprocess the online discussion threads from Reddit
\item Construct a graph of social media posts
\item Perform parameter inference for new discussion threads from the post graph 
\end{enumerate}

The preprocessing step prepares the training data. For each post in the training set, Hawkes model parameters are fit to existing, completed discussion threads. Next, we construct a similarity network, among all posts in the training set, that connects posts based on their titles and authors. These two steps only need to be performed once, after which the fitted training parameters and post graph can be used to simulate any number of unknown posts from the same community.

To simulate a discussion thread for a new post, the model first uses a modified node2vec process to infer simulation parameters from training data history, based on the position of the new post within the established post graph. Simulation parameters are fed into the Hawkes model defined earlier, which uses six parameters to generate a full comment thread, including tree structure and comment timestamps. If we wish to use partial observations of a discussion thread, we also include an optional parameter refitting step before simulating the comment tree.

\nop{
Finally, we simulate the discussion thread discuss accuracy in the proceeding section.
}

\vspace{.2cm}
\noindent\textbf{Step 1: Preprocessing.}
For a given Reddit post, the training data set will contain only posts from the same subreddit; these training posts are generally taken from the time immediately preceding the test set. Formally, for a set of $n$ test posts $p_0, p_1, p_2, ... p_n$, where $p_i$ is the $i^\textrm{th}$ post chronologically and $p_0$ is the first test post, the training set is defined as $p_{-m}, p_{-m+1}, ... p_{-1}$, or the set of $m$ posts immediately preceding $u_0$ chronologically. 

Recall that the Hawkes model used in the present work involves two different distributions. Following the empirical observations of previous work, the Weibull distribution is used to model the dynamics of top-level comments, \ie, comments that are in reply to the post, and produces parameters $a,b, \textrm{ and } \alpha$. The lognormal distribution is used to model the dynamics of comments that are made in reply to other comments and produces parameters $\mu$, and $\sigma$. Finally, recall the branching factor $n_b$, which controls the number of children for each comment~\cite{medvedev2018modelling}.

Together, the six parameters $a, b,\alpha,\mu,\sigma, \textrm{and } n_b$ characterize the structure, size, and temporal dynamics of a complete discussion cascade. For each post in the training set, we fit all six model parameters using the L-BFGS-B optimizer where parameters are constrained to be positive~\cite{zhu1997algorithm}.

\vspace{.2cm}
\noindent\textbf{Step 2: Constructing the Post Graph.}
Next, we create a graph of social media posts as follows. Let $G_{\textrm{sr}}=(V,E)$ be an undirected weighted graph with posts $p\in V$ contributed to a particular subreddit representing the graph's vertices having properties user, title and Hawkes parameters fitted from the previous step $p_a$, $p_b$, $p_\alpha$, $p_\mu$, $p_\sigma$, $p_{n_b}$. 

Weighted edges between posts $(u, v, w)\in E$ are constructed in the following way. Popular users tend to post popular content\cite{khamis2017self}, so posts submitted by the same user are connected $(u, v, 1)$ $|$ $\forall_{u,v} \textrm{ author}(u) = \textrm{author}(v) \wedge u\ne v$. In addition, we also consider the text of the post title. Each title is tokenized and case-normalized into one or more tokens. Edges are created among posts with overlapping titles and weighted according to their overlap $(u, v, \mathcal{J}(\textrm{title}(u), \textrm{title}(v))$ $|$ $\forall_{u,v} \textrm{ } u\ne v$ where $\mathcal{J} \in [0,1]$ is defined to be the Jaccard coefficient of the tokens in the post's titles. Posts with more similar titles will receive a higher weight and identical titles will receive a weight of 1. Edge weights are additive and will therefore always be between 0 and 2.

Because the number of possible post pairs is very large, $|V|*(|V|-1)$, we limit the scope of the graph by only taking the top-$n$ edges with the largest weight for each node in the graph. Because node2vec only performs $r$ random walks per node, limiting to the top-$n$ edges for $n > r$ does not cause significant information loss. The top-$n$ filter is applied to each node without considering neighbor nodes, ie, each node is guaranteed to have at least the $n$ incident edges with the highest weights included in the graph, but may have more depending on the top-$n$ edges of other nodes. Even in the worst-case, this top-$n$ criteria limits the total edges in the graph to $|V|*n$. 

The post graph is constructed for each subreddit and connects existing training posts. A new post $x$ having a submitting user and a title can be connected into the graph creating $G^*_\textrm{sr}$. This version of the graph, containing all training posts and one new post, is then used to infer parameters to predict the discussion cascade for $x$. A new version of $G^*_\textrm{sr}$, based on the same base graph $G_{\textrm{sr}}$, is created for each new post $x$, so that $G^*_\textrm{sr}$ always contains exactly one unknown post $x$.

\vspace{.2cm}
\noindent\textbf{Step 3: Parameter Inference.}
At this point, $G^*_\textrm{sr}$ consists of many nodes representing social media posts and discussion cascades that have been fit to the Hawkes model discussion earlier. As a result, each node is labeled with parameters $p_a$, $p_b$, $p_\alpha$, $p_\mu$, $p_\sigma$, $p_{n_b}$, except the new post $x$ whose parameters cannot be fitted from the empty or very small discussion cascade. By assuming that similar posts will have similar discussion threads, and by wiring $G^*_\textrm{sr}$ according to that assumption, we can use the new node's position in the graph to infer model parameters for the new post $x$.

An ideal choice for network-based parameter inference can be found in graph-based representation learning algorithms like LINE~\cite{tang2015line}, DeepWalk~\cite{perozzi2014deepwalk}, and node2vec~\cite{grover2016node2vec}. Based on the ``you shall be known by the company you keep'' principle frequently cited in text-based representation learning systems like word2vec~\cite{mikolov2013distributed}, the goal of graph representation learning systems is to learn a vector-representation (called an \emph{embedding}) of each node based on the embeddings of the node's neighborhood. 

Because graphs are easily represented as matrices, there are many classic systems for unsupervised feature learning. Traditional dimensionality reductions like PCA are available, but are computationally intractable on large real-world graphs. Graph representation learning, on the other hand, is iterative, easily distributed and therefore much more scalable. 

Graph representation systems utilize a $d$-dimensional vector where the number of dimensions $d$ is defined by the user. Typically, $d$ is 200-500, and vector values are randomly initialized. Despite these random starting points, after optimization, the feature vectors of any two close neighbors tend to converge to similar values as defined by the objective function. However, a single feature vector, when removed from the context of the set, contains no meaningful information; the vectors can only be used for similarity comparisons between objects, and do not describe or capture the characteristics of the object itself.

Recent work reconsiders graph representation as a matrix factorization problem and unifies DeepWalk, LINE, node2vec, etc. into a matrix factorization task\cite{qiu2018network}. This shared representation reconsiders the differences between the approaches to be differences in the transition probabilities for random walks over the graph. Still, the learned representations are not meaningful outside of their relative distance in a vector-space.

\begin{figure}[t]
    \centering
    \begin{tikzpicture}

\node[shape=circle,draw=black] (u) at (0,2.5) {u};
\node[shape=circle,draw=black] (x) at (1.5,1) {x};
\node[shape=circle,draw=black] (v) at (3,2.5) {v} ;
\node (a) at (-1, 1.5) {};
\node (b) at (4, 3.5) {};
\node (c) at (4, 1.5) {};

\path [-] (u) edge[thick, bend left=10] node[left] {$w_1$} (x);
\path [-] (u) edge[thick, bend left=10] node[above] {$w_2$} (v);
\path [-] (v) edge[thick, bend right=10] node[right] {$w_3$} (x);

\draw[] (u) edge[thick] ($(u)!0.60!(a)$) edge [dashed] ($(u)!0.8!(a)$);
\draw[] (v) edge[thick] ($(v)!0.60!(b)$) edge [dashed] ($(v)!0.8!(b)$);
\draw[] (v) edge[thick] ($(v)!0.60!(c)$) edge [dashed] ($(v)!0.8!(c)$);

\node at (-0.5,3.0) {$u_{ [a, b,  \alpha , \mu, \sigma, n_b]}$};
\node at (4.5,2.5) {$v_{ [a, b,  \alpha , \mu, \sigma, n_b]}$};
\node at (2.0,0.6) {$x_{ [a, b,  \alpha , \mu, \sigma, n_b]}$};

\end{tikzpicture}
    \vspace{-3mm}
    \caption{Post graph with parameter embeddings for node2vec parameter inference. Each node's embedding represents the model parameters $a, b, \alpha, \mu, \sigma, n_b$.}
    \label{fig:param_infer_graph}
\end{figure}
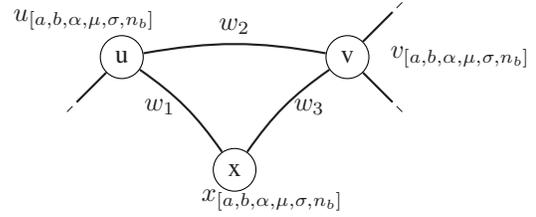

Graph representation learning systems appear to be a convenient solution to our parameter inference problem, but a few modifications are necessary. Instead of initializing the node embeddings to $d=$ 200-500 random values, we set the embedding dimension $d = 6$ (the number of parameters in the Hawkes model) and initialize the embeddings to the fitted parameter values. In other words, the embedding of each node $u\in V$ is $u_{[a,b,\alpha,\mu,\sigma,n_b]}$, which is set to the Hawkes parameters fitted based on $u$'s discussion cascade. The graph snippet in Fig.~\ref{fig:param_infer_graph} illustrates the embeddings for each node. Unlike in node2vec, etc. where the embeddings are only meaningful relative to other embeddings, these embeddings actually represent parameters of the Hawkes model. 

A new post $x$, which does not yet have an extensive discussion thread, is given an initial embedding. The initial embeddings of node2vec are random and range from -0.5 to 0.5; however, Hawkes parameters have their own range depending on their function. We set the initial embedding of $x$ to $(1, 2, 0.75, 0.15, 1.5, 0.05)$. These specific values occur frequently in the set of fitted parameters and, like the median cascade in our data set, these values tend to generate very small discussion cascades. 

Stochastic Gradient Descent (SGD) is used to optimize the parameters and, like in node2vec, the gradients are estimated using backpropagation. The learning rate $r$ starts at $0.0001$ and decreases as the number of iterations increases~\cite{perozzi2014deepwalk,grover2016node2vec}.

At this point, the optimizer typically runs and adjusts each node's embedding according to the loss function. Our case is different. The parameters of each training post were fitted previously by the Hawkes process, so these values should be resistant to significant change. And their resistance to change should be based on the performance of the Hawkes-optimizer. If the log-likelihood of the Hawkes-parameters of some post was very high, then there is little reason to update the post's embedding. On the other hand, if a post's Hawkes-parameters were under-fit (or not fit at all in the case of a new post $x$), then it's embedding may benefit from larger updates. We therefore define a node-specific learning rate $r_u$ as follows:

\begin{equation} \label{eq:rate}
r_u = \left(1-g_u\right) r \left(1-\frac{k}{K+1}\right)
\end{equation}

\noindent where $k$ is the current iteration out of $K$ total iterations such that $1-\frac{k}{K+1}$ is a percent of training complete, $r$ is the base learning rate, and $g_u$ is the goodness of fit score from the Hawkes process for node $u$, normalized between [0.45,0.85]. In rare cases where the Hawkes-optimizer failed to converge, $g_u$ is set to 0.95. Let $g_x=0$. The inverse of this normalized goodness of fit score allows the parameters to be adjusted where needed and remain consistent where learning is not needed.

\newcommand{\markerone}{\raisebox{0.5pt}{\tikz{\node[draw,blue!50!black,scale=0.6,regular polygon, regular polygon sides=4,fill=none](){};}}}

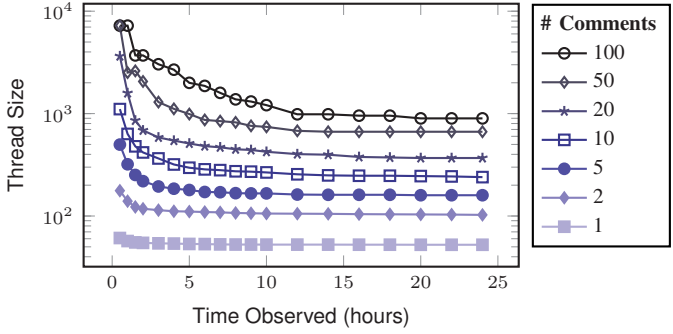
\begin{figure}
    \centering
    \pgfplotstableread{
time	1	2	5	10	20	50	100
0.5	60.97525773	176.7733333	498.1842105	1106.428571	3638	7227	7227
1	56.96401515	139.6220096	319.9753086	633.2631579	1585.916667	2512.666667	7227
1.5	55.23722628	122.6431535	251.7522124	479.0545455	856.6153846	2613.285714	3700.5
2	54.65480427	117.7193676	218.8712121	417.5692308	685.6470588	2059.888889	3700.5
3	54.08978873	114.075188	194.2913907	364.7631579	585.3111111	1295.5	3033.333333
4	53.81260946	111.2747253	184.8024691	318.5934066	544.64	1112.952381	2684.714286
5	53.35590278	110.5018182	179.502994	296.1616162	510.8545455	989.7692308	1997.7
6	53.17647059	109.3417266	171.6514286	285.4174757	481.2033898	867.3548387	1864.181818
7	52.81958763	108.575	170.7443182	280.4666667	468.704918	844.6875	1594.6
8	52.65582192	107.1091549	167.1555556	273.3148148	448.171875	821.1818182	1377.555556
9	52.65582192	106.3741259	166.281768	271.0183486	441.6153846	759.1666667	1311.894737
10	52.56752137	105.6909722	166.281768	266.4144144	423.8382353	742.0810811	1209.428571
12	52.56752137	105.3356401	162.0107527	255.4913793	402.0972222	678.9512195	984.7037037
14	52.56752137	104.9793103	161.171123	249.3193277	396.8767123	665	984.7037037
16	52.47952218	104.2773973	161.171123	247.3416667	377.4285714	665	954.5
18	52.47952218	103.9283276	161.171123	247.3416667	372.8846154	665	954.5
20	52.39182283	103.585034	159.5661376	245.3884298	368.4556962	665	898.1666667
22	52.39182283	103.240678	159.5661376	243.4672131	368.4556962	665	898.1666667
24	52.39182283	102.5589226	159.5661376	239.7580645	368.4556962	665	898.1666667
}{\data}

\begin{tikzpicture}
\begin{axis} [
    thick, 
    width=2.9in,
    height=2.0in, 
    xlabel=\sffamily\footnotesize{{Time Observed (hours)}},
    xlabel near ticks,
    ylabel={\sffamily\footnotesize{Thread Size}},
    ylabel near ticks,
    legend cell align={left},
    legend style={font=\footnotesize},
    legend pos = outer north east,
    ymode=log
]

\addlegendimage{empty legend}
\addplot [mark=o, black] table [x=time, y=100] {\data};
\addplot [mark=diamond, blue!30!black] table [x=time, y=50] {\data};
\addplot [mark=star, blue!60!black] table [x=time, y=20] {\data};
\addplot [mark=square, blue!90!black] table [x=time, y=10] {\data};
\addplot [mark=*, blue!80!white] table [x=time, y=5] {\data};
\addplot [mark=diamond*, blue!50!white] table [x=time, y=2] {\data};
\addplot [mark=square*, blue!30!white] table [x=time, y=1] {\data};

\legend{\hspace{-.7cm}\textbf{\# Comments}, 100, 50, 20, 10, 5, 2, 1}
\end{axis}
\end{tikzpicture}
    \vspace{-6mm}
    \caption{Mean final comment thread size (in number of comments) as a function of observation time for the /r/worldnews/ subreddit. Each line represents a minimum thread size at each point of observation, \eg, the mean final comment thread size for posts with 10 or more observed comments (line indicated by \protect\markerone) in the first half hour is approximately 1,100 comments.}
    \label{fig:observation_time}
\end{figure}

Using the fitted parameter embeddings and the fine-tuned SGD optimizer, the node2vec algorithm is applied to $G_\textrm{sr}^*$, which contains all training posts and one new post. Once node2vec finishes all walks and updates, the final embedding values of the unknown post represent inferred parameters for the Hawkes model. These parameters indicate the predicted comment dynamics of the post, based only on the post title, submitting author, and the subreddit training data. No observed comments are required for this inference step, allowing the CTPM to obtain parameters for new posts.

\subsection*{Partial Fit to Refine Parameters} \label{section_refineparam}

In some cases, it may be useful to wait to gather some data before making a prediction. A post with very few comments after a few hours is unlikely to receive many more. Conversely, a post that garnered lots of immediate attention and has a large comment thread is likely to receive more comments and further attention~\cite{glenski2017rating,cheng2014can}. Fig.~\ref{fig:observation_time} illustrates this dynamic for comment threads in the /r/worldnews/ subreddit. We observe that initial comment velocity is highly predictive of the final comment thread size. For example, if a post in the /r/worldnews/ subreddit receives 5 or more comments in the first half-hour, the final cascade will, on average, consist of 90.53 comments; if the post has fewer than 5 comments in the same period, the final cascade will, on average, only contain 5.07 comments.

When a partial cascade is observed, then we can use this important additional information to improve our prediction of the final discussion cascade. 

In these cases, we apply the same Hawkes parameter fit procedure that was used for training data preprocessing, but slightly modify it to take the inferred parameters into account. First, the optimizer is initialized using the parameters inferred from Step 3. Second, we limit the number of iterations for the Hawkes-optimizer to the number of comments observed. This simple heuristic gives more weight to cascades with more observations. Observed activity is a better predictor of future comment tree growth, but only once there is enough information to perform an accurate fit. Overall, as more comments are observed, the refined parameters approach the `true' fitted parameters of the complete ground-truth cascade.

\subsection*{Cascade Prediction} \label{section_cascadesim}
The Hawkes parameters $a, b, \alpha, \mu, \sigma, n_b$ are fitted with Steps 1-3 above and refined if partial cascades are available. From these parameters we simulate the discussion cascade.

There are three key steps that need to be considered when simulating a discussion cascade. The first is to estimate how many `immigrant' events exist in the cascade. Let $T$ be a predetermined cutoff time $t\in[0,T]$. We simulate immigrant arrival times $\tau<T$ as the cumulative sums of $n$ random variables $\tau_1, \tau_2, \ldots \tau_n$ at a rate of $h(t)$. 

Recall that we modelled immigrant events as a Weibull distribution with parameters $a, b, \alpha$. Thus
$$h(t) = \left( a\frac{\alpha}{b}\right)  \left(\frac{t}{b}\right)^{\alpha-1} \exp\left(-\left(t/b\right)^\alpha\right)$$

\noindent and $\tau$ is sampled from $h(t)$ with the thinning algorithm.

Next, for each $\tau_i<t$ we simulate `offspring' births at a rate of $n_b\phi(t-\tau_i)$. Again recall that we modelled offspring events as a lognormal distribution with parameters $\mu, \sigma$. Thus,

$$\phi(t-\tau_i) = \frac{1}{\sigma (t-\tau_i) \sqrt{2\pi}} \exp\left( -\frac{\left(\log (t-\tau_i) - \mu\right)^2 }{2\sigma^2}\right)$$

\noindent and additional events $\tau$ are sampled from $n_b\sum_{i:t>t_i}\phi\left(t-\tau_i\right)$ with the Thinning algorithm.

Finally, each offspring event may generate additional offspring. So this process iterates until $t>T$ or $n>N$, where $N$ is the maximum number of events allowed. For $n_b<1$ this process will likely die out on its own.

If there exists a partial cascade, then we initialize values for $\tau$ according to the observed data and set $t$ to the time of the most recent comment, or the last time observed. In this case, additional root-level comments may be simulated, and comment replies are generated for both observed and simulated comments.

This produces a simulation of a full discussion thread, including comment timestamps and a tree structure (as illustrated in Fig.~\ref{fig:hawkes_tree}). Note that the simulation does not generate text content or assign comment-authors. 

\section{Methodology}

In this section, we describe the training and testing procedures used to evaluate the Comment Thread Prediction Model (CTPM) introduced in the present work. To this end, we first introduce a large dataset from Reddit, describe the details of 5 different testing scenarios, and compare our results with 4 other state of the art and baseline models. 

\nop{
\begin{table} 
\centering
\caption{Days covered by train and test sets for each subreddit. Testing data begins on December 1, 2017 and moves forward until 1000 posts are retrieved; training data ends on November 30, 2017 and moves backward in time until 10000 posts are retrieved.}
\label{tbl:post_set_days}
\begin{tabular}{rccccc}
\toprule
                  & test posts & & \multicolumn{3}{c}{train posts} \\
   subreddit      & 1000  & & 1000  & 5000 & 10000 \\
\midrule
aww               & 1    & & 2     & 6    & 10    \\
changemyview      & 22   & & 21    & 110  & 234   \\
explainlikeimfive & 3    & & 3     & 13   & 28    \\
IAmA              & 31   & & 25    & 131  & 249   \\
nottheonion       & 11    & & 10    & 48   & 94    \\
science           & 11    & & 10    & 51   & 97    \\
Showerthoughts    & 1    & & 1     & 4    & 9     \\
sports            & 4    & & 4     & 18   & 39    \\
worldnews         & 2    & & 2     & 7    & 13   \\
\bottomrule
\end{tabular}
\end{table}
}

\vspace{.2cm}
\noindent\textbf{Dataset.}
To evaluate the CTPM, we apply it to real-world posts on Reddit. Nine subreddits of varying size and popularity were selected as the source of data: /r/aww/, /r/changemyview/, /r/explainlikeimfive/, /r/IAmA/, /r/nottheonion/, /r/science/, /r/Showerthoughts/, /r/sports/, and /r/worldnews/. For each subreddit, we test our model on the first 1000 posts of December 2017. The size of the training set varies, but is always taken as the $m$ posts immediately preceding the first test post; in this way, the training data ends at November 2017, and covers as many previous days as are required to reach the desired training set size. 

\nop{
Table \ref{tbl:post_set_days} lists the number of days covered by each testing and training set. For example, in the /r/aww/ subreddit, 1000 posts were captured within 1 day starting at December 1, 2017; training sets of 1,000, 5,000, and 10,000 posts were found by scraping the /r/aww/ backwards in time for 2, 6, and 10 days respectively.
}

As shown in Fig.~\ref{fig:size_dist}, most comment threads receive few comments; in fact, more than 83\% of all comment threads in our test data set have fewer than 10 comments. The distribution does vary by subreddit, however: only 55.6\% of /r/changemyview/ posts have fewer than 10 comments, while 97.3\% of /r/sports/ posts meet this criteria. The frequency of small threads supports the need for a model that can accurately predict all sizes of cascades.

\vspace{.2cm}
\noindent\textbf{Model Comparison.}
To demonstrate the effectiveness of our model, we compare results against four alternative models:

\begin{itemize}
	\item \textbf{Hawkes model of discussion trees \cite{medvedev2018modelling}.} Uses the same parameter fit and tree simulation as our model, but requires at least 10 observed comments to estimate parameters. 
	\item \textbf{Random cascade baseline.} Draws an existing cascade from the training dataset at random. Does not take observed comments into account.
	\item \textbf{Random simulation baseline.} Draws fitted parameters from the training dataset at random and simulates a cascade using these parameters, including observed comments in the simulated thread.
	\item \textbf{Average simulation baseline.} Take the average of all fitted parameters in the training dataset and use these parameters to simulate a cascade, including observed comments in the simulated thread. 
\end{itemize}

The Hawkes model does not make use of any training data, and instead relies solely on observed comments to predict future thread growth. This means that the model is incapable of making a prediction for new posts without comments. In general, we observed that 10 comments are needed before the model will return any fitted parameters; so, we begin the evaluation of this model when the cascade has at least 10 comments. This means the Hawkes model cannot simulate new posts, or posts that never receive 10 comments.

The remaining random and average baseline models utilize the same training data set as our Comment Thread Prediction Model, but only to draw parameters from this set. Observed cascade behavior does not contribute to the prediction, beyond serving as a starting point for the simulated comments. Because most cascades are small, these random predictors are actually formidable comparisons.

\vspace{.2cm}
\noindent\textbf{Evaluation Metrics.}
We evaluate the various models by (1) selecting a test post $u$, (2) using each model to simulate a full cascade $u^\ast$, and (3) comparing the similarity of the actual cascade of $u$ and the $u^\ast$ generated by each model. Since $u$ and $u^\ast$ are both trees, there exist several methods by which we can evaluate their topological distance. We start with simple metrics including cascade size (in number of nodes), depth, and breadth. 

We also compute and compare the \textit{structural virality} of $u$ and $u^\ast$. Also known as the Weiner index, the structural virality is defined as the average distance between all pairs of nodes in a tree \cite{goel2015structural}. Then, $\textrm{vir}(u)$ can be defined as follows for trees with $n > 1$ nodes:

\begin{figure*}
    \centering
    \pgfplotstableread{
subreddit	avg_err_10K	ci_10K	avg_err_5K	ci_5K	avg_err_1K	ci_1K
aww	0.54183949	0.016899431	0.543653585	0.017123698	0.558920985	0.018021442
changemyview	0.830767108	0.015884706	0.819570942	0.014296357	0.856108097	0.017816662
explainlikeimfive	0.668901052	0.021104511	0.632767373	0.016337465	0.637130661	0.017231197
IAmA	0.697536162	0.014656064	0.704648504	0.015197478	0.65523567	0.010973481
nottheonion	0.652647064	0.02197677	0.723487542	0.026621111	0.721883952	0.026088155
science	0.7030038	0.024975845	0.692122338	0.026252966	0.845488153	0.033962737
Showerthoughts	0.646609305	0.021413051	0.597086244	0.019424624	0.603365494	0.01955799
sports	0.490385696	0.022487703	0.50441234	0.022570547	0.685137441	0.028938571
worldnews	0.519161551	0.015182372	0.523787247	0.014921702	0.734879025	0.026762152

}\datasize

\pgfplotstableread{
subreddit	avg_err_10K	ci_10K	avg_err_5K	ci_5K	avg_err_1K	ci_1K
aww	0.417937941	0.011778107	0.420514546	0.011665113	0.42381142	0.011865209
changemyview	0.685889974	0.010580981	0.689627161	0.0104183	0.696654089	0.01079348
explainlikeimfive	0.454293892	0.010882744	0.461360933	0.010582844	0.459557078	0.010739129
IAmA	0.521093201	0.008736818	0.521795052	0.009081262	0.506317035	0.008505657
nottheonion	0.429494408	0.010072534	0.428444075	0.011063683	0.42315476	0.011441451
science	0.490724041	0.012907924	0.484838102	0.012919929	0.517717973	0.014431537
Showerthoughts	0.463933639	0.012445957	0.443388495	0.012056507	0.437951468	0.012042466
sports	0.352928301	0.013410105	0.360304838	0.013497579	0.448688053	0.014871705
worldnews	0.413828349	0.011179004	0.417845235	0.011087313	0.49763606	0.014153292

}\datadepth

\pgfplotstableread{
subreddit	avg_err_10K	ci_10K	avg_err_5K	ci_5K	avg_err_1K	ci_1K
aww	1.0372	0.046243656	1.0446	0.046306702	1.0472	0.046140168
changemyview	7.1646	0.221285241	7.1674	0.221612526	7.1728	0.220849139
explainlikeimfive	1.6248	0.058670892	1.6284	0.05880358	1.625	0.058210435
IAmA	3.2898	0.143819737	3.278	0.143791154	3.2658	0.14394291
nottheonion	1.7354	0.085738901	1.797	0.086789857	1.7594	0.086517311
science	1.9378	0.094942934	1.9352	0.095226826	1.9932	0.094162976
Showerthoughts	1.2272	0.056373281	1.2208	0.056042958	1.225	0.055965638
sports	0.563	0.055574785	0.5612	0.055446231	0.651	0.056498654
worldnews	1.7292	0.100317487	1.722	0.100122956	1.7804	0.099095034
}\databreadth

\pgfplotstableread{
subreddit	avg_err_10K	ci_10K	avg_err_5K	ci_5K	avg_err_1K	ci_1K
aww	0.261929254	0.006255198	0.273162177	0.006835655	0.270259492	0.006674463
changemyview	0.724931727	0.005657828	0.728136152	0.005669023	0.730156312	0.005515036
explainlikeimfive	0.310638092	0.007392577	0.304707247	0.006960514	0.309405544	0.007116112
IAmA	0.396792914	0.009193389	0.414398648	0.009677229	0.371935575	0.008686788
nottheonion	0.329131945	0.008697051	0.33837527	0.008943033	0.337917643	0.009036002
science	0.357471552	0.008425363	0.362460875	0.00838575	0.385165851	0.008888818
Showerthoughts	0.313809259	0.007077004	0.300099056	0.006862007	0.309504894	0.007145162
sports	0.307086624	0.0078217	0.289804093	0.00740195	0.347176431	0.00837002
worldnews	0.32231031	0.008330471	0.34509653	0.008388817	0.374121227	0.008928759
}\datavir

\begin{tikzpicture}
\begin{groupplot}[group style={group size= 4 by 1, horizontal sep=0.2cm}, 
    width=.32\textwidth,
    height=1.7in,
    ybar,
    symbolic x coords={aww, changemyview, explainlikeimfive, IAmA, nottheonion, science, Showerthoughts, sports,  worldnews},
    xlabel near ticks,
    xtick=data,
    xticklabel style={rotate=90},
    ylabel near ticks,
    ymin=0, ymax=2.2,
    xlabel near ticks,]
\nextgroupplot [title={Size}, ylabel={\sffamily\scriptsize{MRE}},bar width=1pt,] 
    \addplot+ [error bars/.cd, y explicit, y dir=both] table[x=subreddit, y=avg_err_1K, y error plus=ci_1K, y error minus=ci_1K] {\datasize};
    \addplot+ [error bars/.cd, y explicit, y dir=both] table[x=subreddit, y=avg_err_5K, y error plus=ci_5K, y error minus=ci_5K] {\datasize};
    \addplot+ [error bars/.cd, y explicit, y dir=both] table[x=subreddit, y=avg_err_10K, y error plus=ci_10K, y error minus=ci_10K] {\datasize};
    \legend{1K, 5K, 10K}
    
\nextgroupplot [title={Depth}, yticklabels={,,},bar width=1pt,] 
    \addplot+ [error bars/.cd, y explicit, y dir=both] table[x=subreddit, y=avg_err_1K, y error plus=ci_1K, y error minus=ci_1K] {\datadepth};
    \addplot+ [error bars/.cd, y explicit, y dir=both] table[x=subreddit, y=avg_err_5K, y error plus=ci_5K, y error minus=ci_5K] {\datadepth};
    \addplot+ [error bars/.cd, y explicit, y dir=both] table[x=subreddit, y=avg_err_10K, y error plus=ci_10K, y error minus=ci_10K] {\datadepth};
    
\nextgroupplot [title={Breadth}, yticklabels={,,},bar width=1pt,
    legend cell align={left},
    legend style={font=\footnotesize},
    legend pos = north east] 
    \addplot+ [error bars/.cd, y explicit, y dir=both] table[x=subreddit, y=avg_err_1K, y error plus=ci_1K, y error minus=ci_1K] {\databreadth};
    \addplot+ [error bars/.cd, y explicit, y dir=both] table[x=subreddit, y=avg_err_5K, y error plus=ci_5K, y error minus=ci_5K] {\databreadth};
    \addplot+ [error bars/.cd, y explicit, y dir=both] table[x=subreddit, y=avg_err_10K, y error plus=ci_10K, y error minus=ci_10K] {\databreadth};
    
\nextgroupplot [title={Structural Virality}, yticklabels={,,},bar width=1pt,] 
    \addplot+ [error bars/.cd, y explicit, y dir=both] table[x=subreddit, y=avg_err_1K, y error plus=ci_1K, y error minus=ci_1K] {\datavir};
    \addplot+ [error bars/.cd, y explicit, y dir=both] table[x=subreddit, y=avg_err_5K, y error plus=ci_5K, y error minus=ci_5K] {\datavir};
    \addplot+ [error bars/.cd, y explicit, y dir=both] table[x=subreddit, y=avg_err_10K, y error plus=ci_10K, y error minus=ci_10K] {\datavir};    
\end{groupplot}
\end{tikzpicture}
    \vspace{-6mm}
    \caption{Mean relative error (MRE) of cascade size at time observed = 0 hours for the CTPM at three training set sizes - 1000, 5000, and 10000 posts (lower is better). Larger training set sizes indicate training examples further away (temporally) from the test set. Larger training set sizes generally improve performance. }
    \label{fig:train_compare}
\end{figure*}
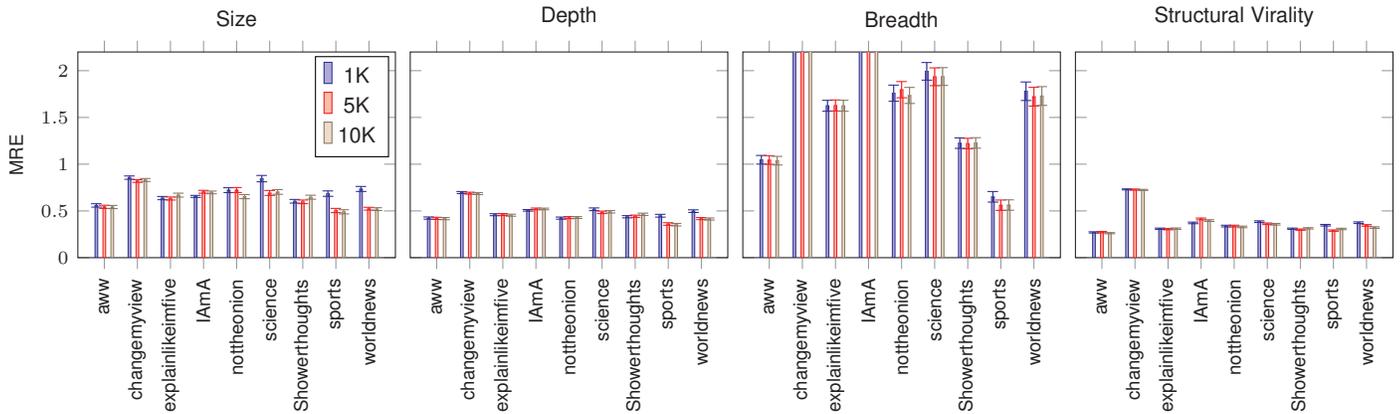

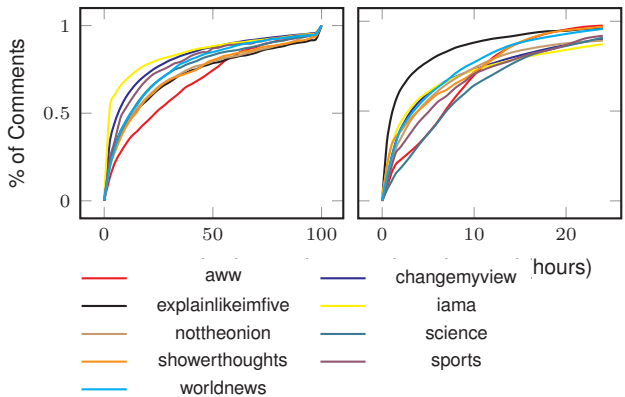
\begin{figure}
    \centering
    \include{./figs/lifetime_vs_comments_mean}
    \vspace{-0.8cm}
    \begin{tikzpicture}
    \begin{customlegend}[ 
    legend columns=2,
  legend style={
    draw=none,
    column sep=2ex,
    font=\sffamily\scriptsize,
  },
  legend entries={\textsf{aww}, \textsf{changemyview}, \textsf{explainlikeimfive}, \textsf{iama}, \textsf{nottheonion}, \textsf{science}, \textsf{showerthoughts}, \textsf{sports}, \textsf{worldnews}},
  ]
    \addlegendimage{red, thick}
    \addlegendimage{blue, thick}
    \addlegendimage{black, thick}
    \addlegendimage{yellow, thick}
    \addlegendimage{brown, thick}
    \addlegendimage{teal, thick}
    \addlegendimage{orange, thick}
    \addlegendimage{violet, thick}
    \addlegendimage{cyan, thick}
    \end{customlegend}
\end{tikzpicture}
    \vspace{-4mm}
    \caption{Percentage of total comments observed over the cascade lifetime (left) and over time (right) for posts with at least 10 comments. Most comments occur within the first few hours.}
    \label{fig:lifespan}
\end{figure}

\begin{equation}
\textrm{vir}(u) = \frac{1}{n(n-1)}\sum_{i=1}^{n}\sum_{j=1}^{n}d_{ij}
\end{equation}

Because the structural virality is only defined for trees with 2 or more nodes, we only compute and compare this metric for threads where both the true and simulated cascades contain at least one comment (since the initial post counts as a node).

For size, depth, breadth, and structural virality comparisons, we plot the mean relative error (MRE) of each metric, which is defined as: 

\begin{equation}
    \textrm{MRE} = \frac{1}{nr}\sum_{i=0}^{n}\sum_{j=0}^{r}\left|\frac{f(u_{i})-f(u^\ast_{ij})}{f(u_{i})}\right|
\end{equation}

\noindent where $n$ is the number of test posts, $r$ is number of repeated simulation runs, $f(x)$ indicates the metric function (size, breadth, depth, or structural virality) applied to the cascade $x$, $u_i$ is a ground-truth cascade, and $u^\ast_{ij}$ is the corresponding simulated cascade for post $u_i$ in test run $j$. Simply put, the MRE measures how closely the predicted cascade matches the actual cascade over $r=5$ runs.

\vspace{.2cm}
\noindent\textbf{The Cascade Timeline.} Discussion threads on Reddit start with the submission of a post, which may (or may not) produce any number of comments. Typically, most comments occur in the first few hours of a post, and nearly all comments occur in the first day. Observation of a cascade can be defined by either (1) the number of comments, or (2) the timespan observed. Fig.~\ref{fig:lifespan} shows that there are subtle differences between these two approaches. The left-hand plot shows the percentage of a completed cascade's comments that are observed during a percentage of that cascade's lifetime; for all subreddits, roughly 75\% of all comments occur in the first half of a thread's total lifetime. The right-hand figure also examines the growth of cascades, but instead defines the observation window using time instead of the percentage of a cascade's lifetime; 14 hours is sufficient to observe 75\% or more of a cascade's comments. In the present work, we choose to capture our results using time (hours) as the independent axis. This decision allows for a more comprehensive, temporal analysis of the cascades.

The discussion threads generated by all models produce timestamps for each simulated comment event. To evaluate these timestamps, we compare the temporal distribution of predicted thread comments against the ground truth. For this, we use the Komolgrov Smirnov Test (KS-Test). The KS-test finds the widest difference between the cumulative distribution functions of the comparative distributions. To gauge how accurately each model captures the temporal dynamics of the comment threads, we perform a KS-test on the ground-truth and simulated comment times of each thread.

\nop{
\begin{figure}
    \centering
    \pgfplotstableread{
subreddit	time_observed	D	p_val	marker_class
aww	24	0.011	0.999957626	2
aww	22	0.0116	0.99986356	2
aww	20	0.012	0.99973036	2
aww	18	0.0134	0.998182169	2
aww	16	0.0164	0.977516788	2
aww	14	0.028	0.525137547	1
aww	12	0.0388	0.159136546	1
aww	10	0.0552	0.011927505	0
aww	9	0.076	0.000121418	0
aww	8	0.099	1.40E-07	0
aww	7	0.1128	1.03E-09	0
aww	6	0.1252	7.20E-12	0
aww	5	0.1444	1.20E-15	0
aww	4	0.1608	2.66E-19	0
aww	3	0.1864	8.62E-26	0
aww	2	0.223	9.92E-37	0
aww	1.5	0.2468	6.82E-45	0
aww	1	0.2824	1.20E-58	0
aww	0.5	0.3494	1.50E-89	0
aww	0	0.4726	1.76E-163	0

}{\aww}

\pgfplotstableread{
subreddit	time_observed	D	p_val	marker_class
changemyview	24	0.1442	1.32E-15	0
changemyview	22	0.162	1.39E-19	0
changemyview	20	0.1828	8.05E-25	0
changemyview	18	0.205	4.18E-31	0
changemyview	16	0.2244	3.46E-37	0
changemyview	14	0.2602	7.48E-50	0
changemyview	12	0.2904	5.43E-62	0
changemyview	10	0.3242	3.71E-77	0
changemyview	9	0.3532	1.69E-91	0
changemyview	8	0.3732	4.18E-102	0
changemyview	7	0.393	3.52E-113	0
changemyview	6	0.429	8.76E-135	0
changemyview	5	0.4572	4.99E-153	0
changemyview	4	0.4944	7.16E-179	0
changemyview	3	0.5334	3.90E-208	0
changemyview	2	0.5692	5.95E-237	0
changemyview	1.5	0.584	2.06E-249	0
changemyview	1	0.6072	1.38E-269	0
changemyview	0.5	0.6296	8.19E-290	0
changemyview	0	0.6728	0	0

}{\changemyview}

\pgfplotstableread{
subreddit	time_observed	D	p_val	marker_class
explainlikeimfive	24	0.0232	0.756230921	1
explainlikeimfive	22	0.0278	0.534478512	1
explainlikeimfive	20	0.0312	0.38650742	1
explainlikeimfive	18	0.0376	0.18559732	1
explainlikeimfive	16	0.0416	0.109036187	1
explainlikeimfive	14	0.0506	0.027030095	0
explainlikeimfive	12	0.0556	0.011071182	0
explainlikeimfive	10	0.065	0.001646586	0
explainlikeimfive	9	0.061	0.003841738	0
explainlikeimfive	8	0.0722	0.000312907	0
explainlikeimfive	7	0.0848	1.13E-05	0
explainlikeimfive	6	0.089	3.30E-06	0
explainlikeimfive	5	0.0956	4.25E-07	0
explainlikeimfive	4	0.1074	7.59E-09	0
explainlikeimfive	3	0.1286	1.69E-12	0
explainlikeimfive	2	0.1576	1.47E-18	0
explainlikeimfive	1.5	0.175	8.78E-23	0
explainlikeimfive	1	0.2012	5.59E-30	0
explainlikeimfive	0.5	0.2658	5.29E-52	0
explainlikeimfive	0	0.4728	1.28E-163	0

}{\explainlikeimfive}

\pgfplotstableread{
subreddit	time_observed	D	p_val	marker_class
IAmA	24	0.1538	1.08E-17	0
IAmA	22	0.1624	1.11E-19	0
IAmA	20	0.1696	2.00E-21	0
IAmA	18	0.1732	2.52E-22	0
IAmA	16	0.1818	1.49E-24	0
IAmA	14	0.1904	6.84E-27	0
IAmA	12	0.1956	2.34E-28	0
IAmA	10	0.2068	1.20E-31	0
IAmA	9	0.2188	2.24E-35	0
IAmA	8	0.2266	6.53E-38	0
IAmA	7	0.2412	6.74E-43	0
IAmA	6	0.2522	7.35E-47	0
IAmA	5	0.2696	1.73E-53	0
IAmA	4	0.2818	2.12E-58	0
IAmA	3	0.2984	1.98E-65	0
IAmA	2	0.3304	4.04E-80	0
IAmA	1.5	0.3582	4.27E-94	0
IAmA	1	0.3832	1.26E-107	0
IAmA	0.5	0.4248	3.63E-132	0
IAmA	0	0.4784	1.65E-167	0

}{\IAmA}

\pgfplotstableread{
subreddit	time_observed	D	p_val	marker_class
nottheonion	24	0.0286	0.497520891	1
nottheonion	22	0.0318	0.363179841	1
nottheonion	20	0.0336	0.298793879	1
nottheonion	18	0.033	0.319313002	1
nottheonion	16	0.0378	0.180962296	1
nottheonion	14	0.0378	0.180962296	1
nottheonion	12	0.0392	0.151016499	1
nottheonion	10	0.0434	0.084316279	1
nottheonion	9	0.0468	0.050355447	1
nottheonion	8	0.0516	0.022763408	0
nottheonion	7	0.065	0.001646586	0
nottheonion	6	0.069	0.000668774	0
nottheonion	5	0.087	5.96E-06	0
nottheonion	4	0.0954	4.54E-07	0
nottheonion	3	0.1126	1.11E-09	0
nottheonion	2	0.1524	2.21E-17	0
nottheonion	1.5	0.1822	1.16E-24	0
nottheonion	1	0.2188	2.24E-35	0
nottheonion	0.5	0.2852	8.31E-60	0
nottheonion	0	0.3708	8.41E-101	0

}{\nottheonion}

\pgfplotstableread{
subreddit	time_observed	D	p_val	marker_class
science	24	0.0306	0.410737295	1
science	22	0.0456	0.060672067	1
science	20	0.0522	0.02050077	0
science	18	0.0622	0.002996385	0
science	16	0.072	0.000328451	0
science	14	0.0854	9.48E-06	0
science	12	0.094	7.09E-07	0
science	10	0.108	6.11E-09	0
science	9	0.1148	4.78E-10	0
science	8	0.1208	4.44E-11	0
science	7	0.1358	6.88E-14	0
science	6	0.1532	1.47E-17	0
science	5	0.17	1.59E-21	0
science	4	0.1868	6.71E-26	0
science	3	0.221	4.41E-36	0
science	2	0.2488	1.29E-45	0
science	1.5	0.2712	4.04E-54	0
science	1	0.2978	3.61E-65	0
science	0.5	0.3252	1.24E-77	0
science	0	0.3956	1.12E-114	0

}{\science}

\pgfplotstableread{
subreddit	time_observed	D	p_val	marker_class
Showerthoughts	24	0.0112	0.999936063	2
Showerthoughts	22	0.0118	0.99980647	2
Showerthoughts	20	0.0136	0.997726117	2
Showerthoughts	18	0.0168	0.971626549	2
Showerthoughts	16	0.0214	0.836153809	1
Showerthoughts	14	0.0282	0.515862005	1
Showerthoughts	12	0.0454	0.062556792	1
Showerthoughts	10	0.0652	0.001576064	0
Showerthoughts	9	0.078	7.23E-05	0
Showerthoughts	8	0.0882	4.19E-06	0
Showerthoughts	7	0.0988	1.50E-07	0
Showerthoughts	6	0.107	8.76E-09	0
Showerthoughts	5	0.1136	7.58E-10	0
Showerthoughts	4	0.1338	1.70E-13	0
Showerthoughts	3	0.157	2.02E-18	0
Showerthoughts	2	0.178	1.48E-23	0
Showerthoughts	1.5	0.1884	2.45E-26	0
Showerthoughts	1	0.2212	3.80E-36	0
Showerthoughts	0.5	0.303	1.89E-67	0
Showerthoughts	0	0.4172	1.71E-127	0

}{\Showerthoughts}

\pgfplotstableread{
subreddit	time_observed	D	p_val	marker_class
sports	24	0.022	0.810627102	1
sports	22	0.0234	0.746831531	1
sports	20	0.0274	0.553341819	1
sports	18	0.0322	0.348140886	1
sports	16	0.0342	0.279219356	1
sports	14	0.0364	0.215378868	1
sports	12	0.0364	0.215378868	1
sports	10	0.0424	0.097393906	1
sports	9	0.0454	0.062556792	1
sports	8	0.0464	0.053611893	1
sports	7	0.0488	0.036514545	0
sports	6	0.0488	0.036514545	0
sports	5	0.051	0.025245193	0
sports	4	0.0544	0.013821654	0
sports	3	0.0644	0.00187614	0
sports	2	0.07	0.000529423	0
sports	1.5	0.073	0.000257403	0
sports	1	0.0776	8.03E-05	0
sports	0.5	0.0888	3.50E-06	0
sports	0	0.1386	1.89E-14	0

}{\sports}

\pgfplotstableread{
subreddit	time_observed	D	p_val	marker_class
worldnews	24	0.0074	1	2
worldnews	22	0.0112	0.999936063	2
worldnews	20	0.012	0.99973036	2
worldnews	18	0.0176	0.95700979	2
worldnews	16	0.0224	0.792932891	1
worldnews	14	0.0296	0.453032877	1
worldnews	12	0.04	0.135775023	1
worldnews	10	0.0526	0.019105865	0
worldnews	9	0.0588	0.005983271	0
worldnews	8	0.0678	0.0008813	0
worldnews	7	0.0792	5.27E-05	0
worldnews	6	0.0936	8.04E-07	0
worldnews	5	0.1064	1.09E-08	0
worldnews	4	0.1258	5.59E-12	0
worldnews	3	0.1452	8.12E-16	0
worldnews	2	0.1654	2.13E-20	0
worldnews	1.5	0.1848	2.34E-25	0
worldnews	1	0.227	4.81E-38	0
worldnews	0.5	0.2806	6.60E-58	0
worldnews	0	0.3384	5.01E-84	0

}{\worldnews}

\begin{tikzpicture}
\begin{axis} [
    thick, 
    width=2.9in,
    height=2.1in, 
    xlabel=\sffont\footnotesize{{Time Observed (hours)}},
    xlabel near ticks,
    ylabel={\sffont\footnotesize{D}},
    ylabel near ticks,
    legend cell align={left},
    legend style={font=\footnotesize},
    legend pos = north east,
    ymode=log,
    cycle list name=color list,
]


\pgfplotsinvokeforeach{\aww, \changemyview, \explainlikeimfive, \IAmA, \nottheonion, \science, \Showerthoughts, \sports, \worldnews}
{
    \addplot+[scatter,
  scatter/classes={0={mark=square},
                   1={mark=o},
                   2={mark=triangle}
                  },
  scatter src=explicit symbolic
  ] table [x=time_observed, y=D, meta=marker_class] {#1};
}

\end{axis}
\end{tikzpicture}
    \caption{Results of KS-test applied to ground truth and simulated cascade lifetimes as observed time increases. Each subreddit evaluated independently.}
    \vspace{-1cm}
    \begin{tikzpicture}
    \begin{customlegend}[ 
    legend columns=2,
  legend style={
    draw=none,
    column sep=2ex,
    font=\sffamily\scriptsize,
  },
  legend entries={\textsf{aww}, \textsf{changemyview}, \textsf{explainlikeimfive}, \textsf{iama}, \textsf{nottheonion}, \textsf{science}, \textsf{showerthoughts}, \textsf{sports}, \textsf{worldnews}},
  ]
    \addlegendimage{red, thick}
    \addlegendimage{blue, thick}
    \addlegendimage{black, thick}
    \addlegendimage{yellow, thick}
    \addlegendimage{brown, thick}
    \addlegendimage{teal, thick}
    \addlegendimage{orange, thick}
    \addlegendimage{violet, thick}
    \addlegendimage{cyan, thick}
    \end{customlegend}
\end{tikzpicture}
    \label{fig:ks_test_subs}
\end{figure}

\begin{figure}
    \centering
    \pgfplotstableread{
model	time_observed	D	p_val	marker_class
model	24	0.0074	1	2
model	22	0.0112	0.999936063	2
model	20	0.012	0.99973036	2
model	18	0.0176	0.95700979	2
model	16	0.0224	0.792932891	1
model	14	0.0296	0.453032877	1
model	12	0.04	0.135775023	1
model	10	0.0526	0.019105865	0
model	9	0.0588	0.005983271	0
model	8	0.0678	0.0008813	0
model	7	0.0792	5.27E-05	0
model	6	0.0936	8.04E-07	0
model	5	0.1064	1.09E-08	0
model	4	0.1258	5.59E-12	0
model	3	0.1452	8.12E-16	0
model	2	0.1654	2.13E-20	0
model	1.5	0.1848	2.34E-25	0
model	1	0.227	4.81E-38	0
model	0.5	0.2806	6.60E-58	0
model	0	0.3384	5.01E-84	0

}{\model}

\pgfplotstableread{
model	time_observed	D	p_val	marker_class
comp	24	0.118	1.37E-10	0
comp	22	0.118	1.37E-10	0
comp	20	0.118	1.37E-10	0
comp	18	0.117	2.03E-10	0
comp	16	0.117	2.03E-10	0
comp	14	0.116	3.00E-10	0
comp	12	0.116	3.00E-10	0
comp	10	0.116	3.00E-10	0
comp	9	0.115	4.43E-10	0
comp	8	0.115	4.43E-10	0
comp	7	0.113	9.53E-10	0
comp	6	0.109	4.24E-09	0
comp	5	0.107	8.76E-09	0
comp	4	0.1226	2.13E-11	0
comp	3	0.15	7.50E-17	0
comp	2	0.1842	3.39E-25	0
comp	1.5	0.215	3.59E-34	0
comp	1	0.2554	4.79E-48	0
comp	0.5	0.313	6.01E-72	0
comp	0	0.469	5.24E-161	0

}{\comp}

\pgfplotstableread{
model	time_observed	D	p_val	marker_class
rand_tree	24	0.0166	0.974685964	2
rand_tree	22	0.0166	0.974685964	2
rand_tree	20	0.0166	0.974685964	2
rand_tree	18	0.0166	0.974685964	2
rand_tree	16	0.0166	0.974685964	2
rand_tree	14	0.0166	0.974685964	2
rand_tree	12	0.0166	0.974685964	2
rand_tree	10	0.0166	0.974685964	2
rand_tree	9	0.0166	0.974685964	2
rand_tree	8	0.0166	0.974685964	2
rand_tree	7	0.0166	0.974685964	2
rand_tree	6	0.0166	0.974685964	2
rand_tree	5	0.0166	0.974685964	2
rand_tree	4	0.0166	0.974685964	2
rand_tree	3	0.0166	0.974685964	2
rand_tree	2	0.0166	0.974685964	2
rand_tree	1.5	0.0166	0.974685964	2
rand_tree	1	0.0166	0.974685964	2
rand_tree	0.5	0.0166	0.974685964	2
rand_tree	0	0.0166	0.974685964	2

}{\randtree}

\pgfplotstableread{
model	time_observed	D	p_val	marker_class
rand_sim	24	0.0462	0.055307087	1
rand_sim	22	0.0464	0.053611893	1
rand_sim	20	0.0454	0.062556792	1
rand_sim	18	0.0518	0.021985684	0
rand_sim	16	0.0524	0.019792363	0
rand_sim	14	0.0514	0.023565473	0
rand_sim	12	0.0548	0.012843152	0
rand_sim	10	0.055	0.012377697	0
rand_sim	9	0.0606	0.00416905	0
rand_sim	8	0.0618	0.003256926	0
rand_sim	7	0.0552	0.011927505	0
rand_sim	6	0.0606	0.00416905	0
rand_sim	5	0.0662	0.001263706	0
rand_sim	4	0.0738	0.000211289	0
rand_sim	3	0.0776	8.03E-05	0
rand_sim	2	0.0974	2.37E-07	0
rand_sim	1.5	0.1024	4.43E-08	0
rand_sim	1	0.1274	2.83E-12	0
rand_sim	0.5	0.1468	3.70E-16	0
rand_sim	0	0.23	4.80E-39	0

}{\randsim}

\pgfplotstableread{
model	time_observed	D	p_val	marker_class
avg_sim	24	0.055	0.012377697	0
avg_sim	22	0.0598	0.004901796	0
avg_sim	20	0.0528	0.018440699	0
avg_sim	18	0.059	0.00575094	0
avg_sim	16	0.0602	0.004521817	0
avg_sim	14	0.0618	0.003256926	0
avg_sim	12	0.0632	0.002426889	0
avg_sim	10	0.1004	8.75E-08	0
avg_sim	9	0.1156	3.51E-10	0
avg_sim	8	0.1418	4.19E-15	0
avg_sim	7	0.1676	6.23E-21	0
avg_sim	6	0.2056	2.76E-31	0
avg_sim	5	0.2572	1.02E-48	0
avg_sim	4	0.325	1.55E-77	0
avg_sim	3	0.3922	1.01E-112	0
avg_sim	2	0.5036	1.42E-185	0
avg_sim	1.5	0.553	1.11E-223	0
avg_sim	1	0.6024	2.39E-265	0
avg_sim	0.5	0.6366	2.77E-296	0
avg_sim	0	0.6468	7.69E-306	0

}{\avgsim}

\begin{tikzpicture}
\begin{axis} [
    thick, 
    width=2.9in,
    height=2.1in, 
    xlabel=\sffont\footnotesize{{Time Observed (hours)}},
    xlabel near ticks,
    ylabel={\sffont\footnotesize{D}},
    ylabel near ticks,
    ylabel style={align=center},
    legend cell align={left},
    legend style={
    draw=none,
    column sep=2ex,
    font=\textsf\tiny,
    nodes={scale=0.6, transform shape},
  }
    legend pos = north east,
    ymode=log,
    legend entries={\textsf{AvgSim}, \textsf{RandSim}, \textsf{RandCascade}, \textsf{Hawkes}, \textsf{CTPM (our model)}}
]

    \addplot+ [green, mark=diamond] table [x=time_observed, y=D, meta=marker_class] {\avgsim};
    \addplot+ [mark=square] table [x=time_observed, y=D, meta=marker_class] {\randsim};
    \addplot [mark=none] table [x=time_observed, y=D, meta=marker_class] {\randtree};
    \addplot+ [] table [x=time_observed, y=D, meta=marker_class] {\comp};
    \addplot+ [mark=o] table [x=time_observed, y=D, meta=marker_class] {\model};

\end{axis}
\end{tikzpicture}
    \caption{Results of KS-test applied to ground truth and simulated cascade lifetimes as observed time increases for posts from /r/worldnews/.}
    \label{fig:ks_test_worldnews}
\end{figure}

\begin{figure}
    \centering
    \include{./figs/lifetime_vs_comments_scatter}
    \caption{Scatter plot of cascade lifetime versus total number of comments for 1000 test post set for each subreddit. \rk{not sure how to make this better... commented out most of the subs for now, since they make compiling take forever}}
    \label{fig:lifetime_vs_comments_scatter}
\end{figure}
}

\nop{
\begin{table*}
\begin{center}
    \caption{Min and max fitted parameters for 1000-post training sets for each subreddit.}
    \label{tbl:min_max_params}
    \begin{filecontents*}{tables/min_max_params.csv}
subreddit,a_min,a_max,b_min,b_max,alpha_min,alpha_max,mu_min,mu_max,sigma_min,sigma_max,n_b_min,_max
aww,0.0001,576.0007887,1,1777.464976,0.15,2.462328935,0.0001,9.469179107,0.0001,3.827781146,0,0.909090909
changemyview,1,123.0152592,1,5408.563035,0.0001,4.589398168,0.0001,7.81800805,0.0001,3.985140433,0,0.97762073
explainlikeimfive,1,252.0058332,1,1098.851996,0.15,1.973689427,0.0001,12.10073165,0.0001,3.899294098,0,0.91356674
IAmA,0.0001,2371.949667,0.0001,3215.860804,0.0001,3.468589627,0.0001,9.976402642,0.0001,3.469918305,0,0.954663887
nottheonion,0.0001,975.1500253,1,6531.904938,0.15,4.376478311,0.0001,8.193446667,0.0001,3.417126499,0,0.943521595
science,0.0001,3551.265652,1,5347.848516,0.15,3.910775016,0.0001,8.614613988,0.0001,3.813200301,0,0.925925926
Showerthoughts,1,212.0001763,1,1241.958057,0.15,1.687321331,0.0001,12.44858763,0.0001,6.495665985,0,0.913043478
sports,1,920.0796824,1,2702.081094,0.15,1.984350827,0.032827046,7.89580216,0.0001,3.80510383,0,0.818181818
worldnews,0.0001,1797.449173,1,3653.038169,0.15,3.229670979,0.15,9.130368339,0.0001,3.308645268,0,0.953367876
\end{filecontents*}
\pgfplotstabletypeset
    [
        multicolumn names,
        col sep=comma, 
        header=true,
        precision=4,
        fixed zerofill,
        fixed,
        set thousands separator={},
        every head row/.style={after row={\toprule & \multicolumn{2}{c}{$a$}& \multicolumn{2}{c}{$b$} & \multicolumn{2}{c}{$\alpha$} & \multicolumn{2}{c}{$\mu$} & \multicolumn{2}{c}{$\sigma$} & \multicolumn{2}{c}{$n_b$}\\ subreddit & min & max & min & max & min & max & min & max & min & max & min & max \\ \midrule}},
        every last row/.style={after row=\bottomrule},
        display columns/0/.style={column name=, column type = {r}, string type},
        display columns/1/.style={column name=, column type = {r}},
        display columns/2/.style={column name=, column type = {r}},
        display columns/3/.style={column name=, column type = {r}},
        display columns/4/.style={column name=, column type = {r}},
        display columns/5/.style={column name=, column type = {r}},
        display columns/6/.style={column name=, column type = {r}},
        display columns/7/.style={column name=, column type = {r}},
        display columns/8/.style={column name=, column type = {r}},
        display columns/9/.style={column name=, column type = {r}},
        display columns/10/.style={column name=, column type = {r}},
        display columns/11/.style={column name=, column type = {r}},
        display columns/12/.style={column name=, column type = {r}},
    ]{tables/min_max_params.csv}
\end{center}
\end{table*}

\begin{table}
\begin{center}
    \caption{Mean fitted parameters for 1000-post training sets for each subreddit.}
    \label{tbl:mean_params}
    \begin{filecontents*}{tables/mean_params.csv}
subreddit,a,b,alpha,mu,sigma,n_b
aww,5.31135278,39.99513943,0.490242786,1.000898031,1.371455615,0.099799762
changemyview,10.46312657,226.4620845,0.55502504,2.374861154,1.64787385,0.457711743
explainlikeimfive,2.999615598,30.43969982,0.640983083,1.254242571,1.265637502,0.168766658
IAmA,29.2644411,46.50874073,0.749402491,1.428431035,1.474819231,0.196907011
nottheonion,8.385983447,64.42562199,0.727996202,0.965395347,1.40963918,0.113952615
science,13.63269791,94.40101409,0.594488413,1.345033152,1.413778218,0.152227577
Showerthoughts,2.789009552,27.26241138,0.480804132,0.954645319,1.353123884,0.123813247
sports,4.647641612,25.78409052,0.266093742,0.462271731,1.491916305,0.037151543
worldnews,8.378395834,47.11944044,0.503352769,0.975274376,1.429953599,0.125265291
\end{filecontents*}
\pgfplotstabletypeset
    [
        col sep=comma, 
        header=true,
        precision=2,
        fixed zerofill,
        fixed,
        set thousands separator={},
        every head row/.style={after row={\toprule subreddit & $a$ &  $b$ & $\alpha$ & $\mu$ & $\sigma$ & $n_b$\\ \midrule}},
        every last row/.style={after row=\bottomrule},
        display columns/0/.style={column name=, column type = {r}, string type},
        display columns/1/.style={column name=, column type = {r}},
        display columns/2/.style={column name=, column type = {r}},
        display columns/3/.style={column name=, column type = {r}},
        display columns/4/.style={column name=, column type = {r}},
        display columns/5/.style={column name=, column type = {r}},
        display columns/6/.style={column name=, column type = {r}},
        display columns/7/.style={column name=, column type = {r}},
    ]{tables/mean_params.csv}
\end{center}
\end{table}
}


\begin{figure*}
    \centering
    \include{./figs/model_comp_by_sub}
    \vspace{-0.4cm}
    \begin{tikzpicture}
    \begin{customlegend}[ 
    legend columns=5,
  legend style={
    draw=none,
    column sep=2ex,
    font=\textsf\tiny,
    nodes={scale=0.8, transform shape},
  },
  legend entries={\textsf{AvgSim}, \textsf{RandSim}, \textsf{RandCascade}, \textsf{Hawkes}, \textsf{CTPM (our model)}},
  ]
    \addlegendimage{green, mark=diamond, thick}
    \addlegendimage{red, mark=square, thick}
    \addlegendimage{black, mark=none, thick}
    \addlegendimage{black, mark=star, thick}
    \addlegendimage{blue, mark=o, thick}
    \end{customlegend}
\end{tikzpicture}
    \vspace{-3mm}
    \caption{Mean relative error (MRE) on a log scale as a function of the time observed. Error bars represent 95\% confidence interval. Lower is better. The CTPM described in the present work outperforms existing methods by a significant margin, especially when the observation time is small.}
    \label{fig:model_comp_by_sub}
\end{figure*}
\begin{figure*}
    \centering
    \input{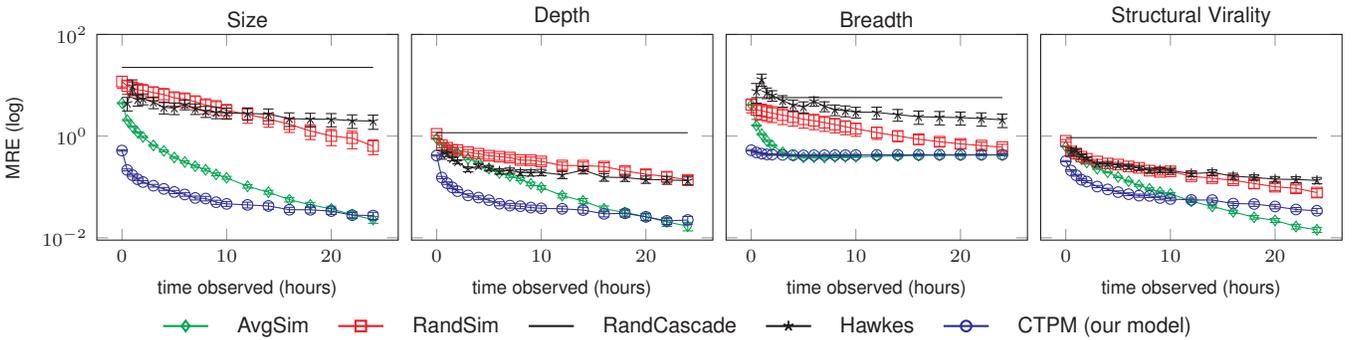}
    \vspace{-0.4cm}
    \begin{tikzpicture}
    \begin{customlegend}[ 
    legend columns=5,
  legend style={
    draw=none,
    column sep=2ex,
    font=\textsf\tiny,
    nodes={scale=0.8, transform shape},
  },
  legend entries={\textsf{AvgSim}, \textsf{RandSim}, \textsf{RandCascade}, \textsf{Hawkes}, \textsf{CTPM (our model)}},
  ]
    \addlegendimage{green, mark=diamond, thick}
    \addlegendimage{red, mark=square, thick}
    \addlegendimage{black, mark=none, thick}
    \addlegendimage{black, mark=star, thick}
    \addlegendimage{blue, mark=o, thick}
    \end{customlegend}
\end{tikzpicture}
    \vspace{-6mm}
    \caption{Mean relative error (MRE) of cascade size, depth, breadth, and structural virality on a log scale for the /r/worldnews/ subreddit as a function of the time observed. Error bars represent 95\% confidence interval. Lower is better. The CTPM described in the present work outperforms existing methods by a significant margin, especially when the observation time is small.}
    \label{fig:worldnews_model_comp}
    \vspace{0.3cm}
\end{figure*}

\section{Results and Discussion}
This section illustrates and discusses the results of our experiments comparing our Comment Thread Prediction Model (CTPM) against existing work and baselines.

\vspace{.2cm}
\noindent\textbf{Size of the Training Set.} We begin with an evaluation of the various training set sizes. We created three training datasets for each subreddit with 1,000, 5,000, and 10,000 posts, all ending on November 30, 2017. The test set begins immediately after. In this scenario there are two competing dynamics: (1) training size and (2) recency. More training data is typically associated with better results, as more data usually provides better generalizability. On the other hand, in this temporal scenario, more data means more \textit{older} data. As we increase the training set size, the additional data items are further and further away from the test set, and may therefore not reflect the current dynamics of the subreddit.

Our first experiment explores this trade-off by comparing the mean relative error (MRE) for the four different cascade metrics in each of the nine subreddits. The results illustrated in Fig.~\ref{fig:train_compare} show that the MRE generally improves as the training set gets bigger, but the effect is small. We therefore use the 10,000 post training set for all remaining tasks.

\vspace{.2cm}
\noindent\textbf{Effect of Observation Time.} As the cascade observation time increases, the amount of information known about the final size and shape of the comment cascade also increases. Recall that most comments occur in the first few hours after submission (Fig.~\ref{fig:lifespan}) and that early comment velocity is very indicative of final comment size (Fig.~\ref{fig:observation_time}). So it is important for a model to perform accurate predictions with as little time elapsed as possible.

The original Hawkes model requires at least 10 comments before any attempt to generate a thread can be made. The CTPM overcomes this via parameter inference on the post graph, allowing for cascade simulation given no observed comments. Here we ask: how does the CTPM compare with Hawkes and other baselines models, especially with little or no observation time?

Results showing the MRE of the comment thread size are illustrated in Fig.~\ref{fig:model_comp_by_sub} for individual subreddits. Here the Hawkes model was not run on threads with fewer than 10 comments. In larger threads the Hawkes parameter estimation still sometimes failed; these failed runs were also removed from results.

We find that the CTPM does indeed outperform the other models, especially at hour 0 when no comments are present. As expected, as time elapses, and more comment information is available, the models tend to improve. The Hawkes model, however, does not perform well even as the comment thread concludes. This lack of improvement in the Hawkes model is probably because of the abundance of small comment threads, which the Hawkes model has difficulty with.

We also observe a significant improvement in the performance of the CTPM after the first hour. An improvement was expected, but the amount of improvement reinforces our previous finding that the first hour in a comment thread's lifecycle is extremely important to its future size, and therefore has enormous predictive power.

\vspace{.2cm}
\noindent\textbf{Metric Comparison.} The previous results in Fig.~\ref{fig:model_comp_by_sub} show only the MRE comparing the comment thread \textit{size} for eight of the nine total subreddits. Using just the /r/worldnews/ subreddit, we also compare the depth, breadth, and structural virality metrics in Fig~\ref{fig:worldnews_model_comp}.

\begin{figure*}
    \centering
    \include{./figs/IAmA_comp_size_classes}
    \vspace{-0.4cm}
    \begin{tikzpicture}
    \begin{customlegend}[ 
    legend columns=5,
  legend style={
    draw=none,
    column sep=2ex,
    font=\textsf\tiny,
    nodes={scale=0.8, transform shape},
  },
  legend entries={\textsf{AvgSim}, \textsf{RandSim}, \textsf{RandCascade}, \textsf{Hawkes}, \textsf{CTPM (our model)}},
  ]
    \addlegendimage{green, mark=diamond, thick}
    \addlegendimage{red, mark=square, thick}
    \addlegendimage{black, mark=none, thick}
    \addlegendimage{black, mark=star, thick}
    \addlegendimage{blue, mark=o, thick}
    \end{customlegend}
\end{tikzpicture}
    \vspace{-3mm}
    \caption{Mean relative error (MRE) of discussion thread size for various final thread sizes on a log scale for the /r/IAmA/ subreddit as a function of the time observed. Error bars represent 95\% confidence interval. Lower is better. The CTPM performs consistently well across various thread sizes. The Hawkes model cannot be computed for small threads ($<$10 comments), so Hawkes results are absent in the left-most panel.}
    \label{fig:IAmA_comp_size_classes}
\end{figure*}
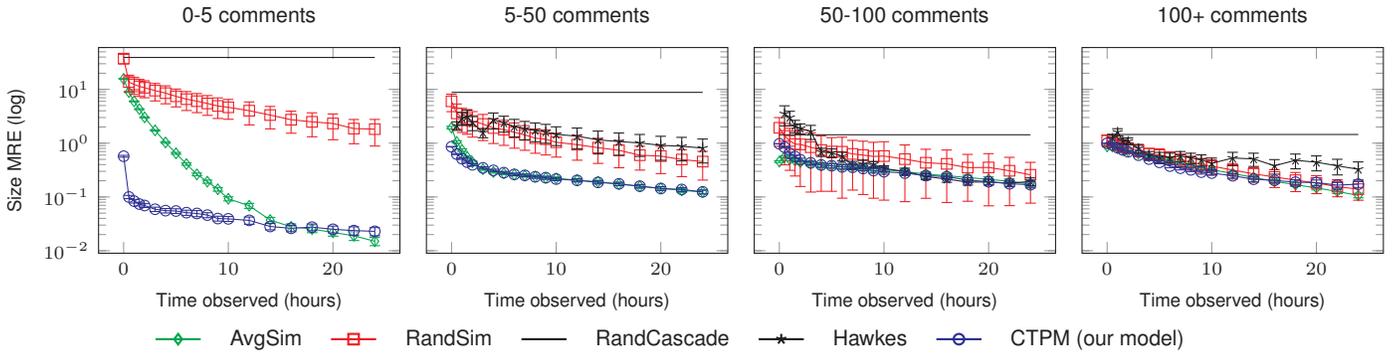

Again we find that CTPM outperforms the other models and baselines, especially when there is little or no observed data. This trend is consistent across all four comparison metrics. 

Interestingly, we find that the AvgSim baseline performs rather well in later stages of the comment thread lifecycle, particularly for the structural virality metric. This is because the majority of cascades are very small (\eg, 809 \mbox{/r/worldnews/} posts have fewer than 5 comments), and the parameters used by the AvgSim baseline tend to generate small cascades. Once the observed portion of the cascade surpasses this typical small size, particularly in the late stages of a comment thread, the AvgSim baseline tends to generate no further comments. The nearly-complete observed tree is then evaluated against the ground truth, yielding small error measurements. CTPM, on the other hand, is not biased towards a particular comment thread size. If a cascade exhibits highly viral behavior early in it's lifetime, the refined parameters will reflect this, causing the model to generate additional comments. These extra comments generated late in the cascade's lifetime allow the less sophisticated AvgSim baseline to overtake CTPM in some performance metrics - particularly structural virality, which is based on the structure of the comment cascade.

\vspace{.2cm}
\noindent\textbf{Comment Thread Size Dynamics.} As we discussed earlier, most comment threads are small. But it is still important to predict both small and large comment threads accurately, and to understand the prediction errors produced by models under various size conditions. 

Fig.~\ref{fig:IAmA_comp_size_classes} shows how model performance varies as the size of the final comment thread changes. In the majority case, \ie, when the final thread has fewer than 5 comments, we observe that CTPM performs significantly better than the other models and baselines, especially when there is little or no time observed in the comment thread's lifecycle. The Hawkes model proposed by Medvedev et al~\cite{medvedev2018modelling}, does not appear in the far-left panel because it could not generate predictions with so few observed. 

As the final comment thread grows larger, the advantage of CTPM decreases. These results indicate that existing models overestimate the comment thread size on the average case. CTPM does not appear to have this overestimation problem; it performs consistently across threads of different sizes.

\vspace{.2cm}
\noindent\textbf{Comment Time Dynamics.} The previous experiments have compared the \textit{final} size and/or shape of the actual comment thread compared to the predicted comment thread. However, these experiments do not consider the temporal aspect of the cascades. Each comment occurs at a specific time and after a specific delay. The Hawkes process used in comparative models, and CTPM, does indeed generate comments at specific times. 

Here we ask: how well do the temporal dynamics of each model compare to the ground truth? To answer this question, we flatten each comment thread into a single timeline of events: the time of each comment relative to the parent post. We compare this linear timeline with the ground truth timeline using the Kolmogorov-Smirnov Test (KS-Test) and report the test statistic. 

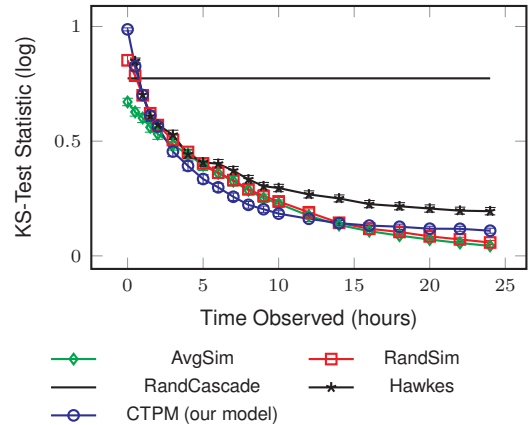
\begin{figure}
    \centering
    \include{./figs/ks_test_models}
    \vspace{-0.4cm}
    \begin{tikzpicture}
    \begin{customlegend}[ 
    legend columns=2,
  legend style={
    draw=none,
    column sep=2ex,
    font=\sffamily\scriptsize,
  },
  legend entries={\textsf{AvgSim}, \textsf{RandSim}, \textsf{RandCascade}, \textsf{Hawkes}, \textsf{CTPM (our model)}},
  ]
    \addlegendimage{green, mark=diamond, thick}
    \addlegendimage{red, mark=square, thick}
    \addlegendimage{black, mark=none, thick}
    \addlegendimage{black, mark=star, thick}
    \addlegendimage{blue, mark=o, thick}
    \end{customlegend}
\end{tikzpicture}    
    \vspace{-0.4cm}
    \caption{Mean results of KS-tests comparing predicted with actual comment thread timestamps as observed time increases for /r/worldnews/ posts with 10 or more comments. Lower is better. The CTPM outperforms the Hawkes model, particularly for large observation windows.}
    \label{fig:ks_test_models}
\end{figure}

The mean KS-Test Statistic of comparing results from the /r/worldnews/ subreddit is illustrated in Fig.~\ref{fig:ks_test_models}. CTPM outperforms the Hawkes model, and performs competitively against the AvgSim and RandSim baselines. As with the structural virality metric results in Fig.~\ref{fig:worldnews_model_comp}, the baseline model overtakes the CTPM when the observed time is high. This is still likely the result of occasional over-simulation by CTPM as compared to the baselines, which are biased towards small threads. Additionally, this KS-Test omits posts with fewer than 10 comments to ensure there are enough events for an accurate assessment. Given that the CTPM vastly outperforms the baseline models for small threads, these results are likely biased against CTPM.

\nop{
\begin{figure}
    \centering
    \pgfplotstableread{
subreddit	avg_err_10K	ci_10K	avg_err_5K	ci_5K	avg_err_1K	ci_1K
aww	0.417937941	0.011778107	0.420514546	0.011665113	0.42381142	0.011865209
changemyview	0.685889974	0.010580981	0.689627161	0.0104183	0.696654089	0.01079348
explainlikeimfive	0.454293892	0.010882744	0.461360933	0.010582844	0.459557078	0.010739129
IAmA	0.521093201	0.008736818	0.521795052	0.009081262	0.506317035	0.008505657
nottheonion	0.429494408	0.010072534	0.428444075	0.011063683	0.42315476	0.011441451
science	0.490724041	0.012907924	0.484838102	0.012919929	0.517717973	0.014431537
Showerthoughts	0.463933639	0.012445957	0.443388495	0.012056507	0.437951468	0.012042466
sports	0.352928301	0.013410105	0.360304838	0.013497579	0.448688053	0.014871705
worldnews	0.413828349	0.011179004	0.417845235	0.011087313	0.49763606	0.014153292

}\data

\begin{tikzpicture}
  \begin{axis}[
    ybar,
    bar width=4.5pt,
    symbolic x coords={aww, changemyview, explainlikeimfive, IAmA, nottheonion, science, Showerthoughts, sports,  worldnews},
    xlabel={subreddit},
    xlabel near ticks,
    xtick=data,
    xticklabel style={rotate=90},
    ylabel={average relative depth error},
    ylabel near ticks,
    legend cell align={left},
    legend style={font=\footnotesize},
    legend pos = north east
  ]
    \addplot+ [error bars/.cd, y explicit, y dir=both] table[x=subreddit, y=avg_err_1K, y error plus=ci_1K, y error minus=ci_1K] {\data};
    \addplot+ [error bars/.cd, y explicit, y dir=both] table[x=subreddit, y=avg_err_5K, y error plus=ci_5K, y error minus=ci_5K] {\data};
    \addplot+ [error bars/.cd, y explicit, y dir=both] table[x=subreddit, y=avg_err_10K, y error plus=ci_10K, y error minus=ci_10K] {\data};
\legend{1K, 5K, 10K}
  \end{axis}
\end{tikzpicture}
    \caption{Average relative cascade depth error at time observed = 0 hours for proposed model at three training set sizes - 1000, 5000, and 10000 posts.}
    \label{fig:train_reldepth_comp_0obs}
\end{figure}

\begin{figure}
    \centering
    \pgfplotstableread{
subreddit	avg_err_10K	ci_10K	avg_err_5K	ci_5K	avg_err_1K	ci_1K
aww	1.0372	0.046243656	1.0446	0.046306702	1.0472	0.046140168
changemyview	7.1646	0.221285241	7.1674	0.221612526	7.1728	0.220849139
explainlikeimfive	1.6248	0.058670892	1.6284	0.05880358	1.625	0.058210435
IAmA	3.2898	0.143819737	3.278	0.143791154	3.2658	0.14394291
nottheonion	1.7354	0.085738901	1.797	0.086789857	1.7594	0.086517311
science	1.9378	0.094942934	1.9352	0.095226826	1.9932	0.094162976
Showerthoughts	1.2272	0.056373281	1.2208	0.056042958	1.225	0.055965638
sports	0.563	0.055574785	0.5612	0.055446231	0.651	0.056498654
worldnews	1.7292	0.100317487	1.722	0.100122956	1.7804	0.099095034

}\data

\begin{tikzpicture}
  \begin{axis}[
    ybar,
    bar width=4.5pt,
    symbolic x coords={aww, changemyview, explainlikeimfive, IAmA, nottheonion, science, Showerthoughts, sports,  worldnews},
    xlabel={subreddit},
    xlabel near ticks,
    xtick=data,
    xticklabel style={rotate=90},
    ylabel={average relative breadth error},
    ylabel near ticks,
    legend cell align={left},
    legend style={font=\footnotesize},
    legend pos = north east
  ]
    \addplot+ [error bars/.cd, y explicit, y dir=both] table[x=subreddit, y=avg_err_1K, y error plus=ci_1K, y error minus=ci_1K] {\data};
    \addplot+ [error bars/.cd, y explicit, y dir=both] table[x=subreddit, y=avg_err_5K, y error plus=ci_5K, y error minus=ci_5K] {\data};
    \addplot+ [error bars/.cd, y explicit, y dir=both] table[x=subreddit, y=avg_err_10K, y error plus=ci_10K, y error minus=ci_10K] {\data};
\legend{1K, 5K, 10K}
  \end{axis}
\end{tikzpicture}
    \caption{Average relative cascade breadth error at time observed = 0 hours for proposed model at three training set sizes - 1000, 5000, and 10000 posts. \rk{should probably combine these three into subfigures, assuming we want them all}}
    \label{fig:train_relbreadth_comp_0obs}
\end{figure}

\begin{figure}
    \centering
    \pgfplotstableread{
subreddit	avg_err_10K	ci_10K	avg_err_5K	ci_5K	avg_err_1K	ci_1K
aww	0.261929254	0.006255198	0.273162177	0.006835655	0.270259492	0.006674463
changemyview	0.724931727	0.005657828	0.728136152	0.005669023	0.730156312	0.005515036
explainlikeimfive	0.310638092	0.007392577	0.304707247	0.006960514	0.309405544	0.007116112
IAmA	0.396792914	0.009193389	0.414398648	0.009677229	0.371935575	0.008686788
nottheonion	0.329131945	0.008697051	0.33837527	0.008943033	0.337917643	0.009036002
science	0.357471552	0.008425363	0.362460875	0.00838575	0.385165851	0.008888818
Showerthoughts	0.313809259	0.007077004	0.300099056	0.006862007	0.309504894	0.007145162
sports	0.307086624	0.0078217	0.289804093	0.00740195	0.347176431	0.00837002
worldnews	0.32231031	0.008330471	0.34509653	0.008388817	0.374121227	0.008928759

}\data

\begin{tikzpicture}
  \begin{axis}[
    ybar,
    bar width=4.5pt,
    symbolic x coords={aww, changemyview, explainlikeimfive, IAmA, nottheonion, science, Showerthoughts, sports,  worldnews},
    xlabel={subreddit},
    xlabel near ticks,
    xtick=data,
    xticklabel style={rotate=90},
    ylabel={average relative structural virality error},
    ylabel near ticks,
    legend cell align={left},
    legend style={font=\footnotesize},
    legend pos = north east
  ]
    \addplot+ [error bars/.cd, y explicit, y dir=both] table[x=subreddit, y=avg_err_1K, y error plus=ci_1K, y error minus=ci_1K] {\data};
    \addplot+ [error bars/.cd, y explicit, y dir=both] table[x=subreddit, y=avg_err_5K, y error plus=ci_5K, y error minus=ci_5K] {\data};
    \addplot+ [error bars/.cd, y explicit, y dir=both] table[x=subreddit, y=avg_err_10K, y error plus=ci_10K, y error minus=ci_10K] {\data};
\legend{1K, 5K, 10K}
  \end{axis}
\end{tikzpicture}
    \caption{Average relative cascade structural virality error at time observed = 0 hours for proposed model at three training set sizes - 1000, 5000, and 10000 posts. Only posts where both true and simulated cascades have at least one comment are considered.}
    \label{fig:train_relstructvir_comp_0obs}
\end{figure}
}

\nop{
\begin{table}
\begin{center}
    \caption{Count of posts in each size class by subreddit. Based on final size of true cascades.}
    \label{tbl:posts_per_size}
    \pgfplotstabletypeset
    [
        col sep=comma, 
        header=true,
        set thousands separator={},
        every head row/.style={after row={\toprule & \multicolumn{4}{c}{number of comments}\\ subreddit & $[0, 5)$ & $[5, 50)$ & $[50, 100)$ & $[100, +\inf)$ \\ \midrule}},
        every last row/.style={before row={\toprule}, after row=\bottomrule},
        display columns/0/.style={column name=, column type = {r}, string type},
        display columns/1/.style={column name=, column type = {r}},
        display columns/2/.style={column name=, column type = {r}},
        display columns/3/.style={column name=, column type = {r}},
        display columns/4/.style={column name=, column type = {r}},
    ]{tables/size_class_counts.csv}
\end{center}
\end{table}

\begin{figure}
    \centering
    \pgfplotstableread{
time	0-5_err	5-50_err	50-100_err	100+_err	0-5_ci	5-50_ci	50-100_ci	100+_ci
0	0.571706667	0.858558583	0.97435149	0.993685891	0.02200783	0.008772891	0.002517974	0.000835913
0.5	0.099978667	0.626909833	0.841065754	0.929877881	0.00700833	0.01365265	0.014022534	0.008061461
1	0.083589333	0.514455934	0.64706769	0.824305455	0.006593068	0.015952416	0.029334945	0.019482927
1.5	0.073861333	0.443740736	0.563836657	0.749243589	0.006013952	0.016936318	0.032325251	0.024264135
2	0.069392	0.396774972	0.491431424	0.691933673	0.00566444	0.016803103	0.032772027	0.026854961
3	0.059114667	0.347298119	0.431516845	0.597998156	0.005476213	0.016757051	0.034087736	0.02732824
4	0.055685333	0.31209519	0.393478992	0.506573664	0.00526681	0.016061859	0.035852617	0.027030745
5	0.054773333	0.282988274	0.380199002	0.440593113	0.005492255	0.015495382	0.034905648	0.025731221
6	0.051098667	0.262764705	0.36188385	0.375093053	0.005595843	0.014901255	0.034425064	0.024711616
7	0.049349333	0.251930287	0.355916706	0.342278646	0.00642695	0.014856015	0.032954569	0.023569031
8	0.046005333	0.237217021	0.337898649	0.319337998	0.004930155	0.014660719	0.031449536	0.0231116
9	0.039717333	0.228306408	0.309204549	0.289459682	0.004089149	0.014271795	0.030583909	0.023244684
10	0.039173333	0.215556875	0.295088037	0.27800541	0.004150428	0.014136857	0.030893794	0.024395722
12	0.036693333	0.203687847	0.277865065	0.247983698	0.005658047	0.013643758	0.02744237	0.023890036
14	0.028272	0.186608747	0.249880743	0.213615905	0.003829886	0.012624783	0.02757826	0.022790707
16	0.026133333	0.175581715	0.238166372	0.203467987	0.003552081	0.012524988	0.02689897	0.022697371
18	0.027466667	0.158609581	0.206325573	0.188129844	0.00388259	0.011941373	0.02274671	0.020390122
20	0.025072	0.143736757	0.192108356	0.185161812	0.003708011	0.011516997	0.022519141	0.021834957
22	0.023701333	0.137817873	0.175812186	0.166169737	0.004163767	0.011755856	0.021215196	0.019620563
24	0.022970667	0.12440823	0.16908597	0.172826051	0.003973444	0.010535204	0.021824884	0.022339177

}{\data}

\begin{tikzpicture}
\begin{axis} [
    thick, 
    xlabel={time observed (hours)},
    xlabel near ticks,
    ylabel={average relative size error},
    ylabel near ticks,
    legend cell align={left},
    legend style={font=\footnotesize},
    legend pos = south east
]

\addplot+ [error bars/.cd, y explicit, y dir=both] table [x=time, y=0-5_err, y error plus=0-5_ci, y error minus=0-5_ci] {\data};
\addplot+ [error bars/.cd, y explicit, y dir=both] table [x=time, y=5-50_err, y error plus=5-50_ci, y error minus=5-50_ci] {\data};
\addplot+ [error bars/.cd, y explicit, y dir=both] table [x=time, y=50-100_err, y error plus=50-100_ci, y error minus=50-100_ci] {\data};
\addplot+ [error bars/.cd, y explicit, y dir=both] table [x=time, y=100+_err, y error plus=100+_ci, y error minus=100+_ci] {\data};

\legend{0-5, 5-50, 50-100, 100+}
\end{axis}
\end{tikzpicture}
    \caption{Average relative size error for different /r/IAmA/ cascade sizes as observed time increases. \rk{make this pretty} \rk{include other models here? would maybe need to break size classes into separate plots then, but might be worth it}}
    \label{fig:IAmA_size_classes}
\end{figure}
}

\nop{
\section{Related Work}

\subsection{Hawkes Processes for Popularity Prediction}

\cite{medvedev2018modelling} predicting Reddit cascades (including tree structure) from some number of observed comments (defined by time) - this is what our fit/sim code/process is based on, but we added the infer for 0 observed comments

\cite{zhao2015seismic} predict final retweet counts via self-exciting point process based on reshare history (defined by time) so far

\cite{kobayashi2016tideh} twitter prediction based on observation period, takes circadian rhythms into account, relies on observed retweets and social network structure, Hawkes for prediction

\cite{rizoiu2017tutorial} general Hawkes info, retweet prediction example

\cite{zhou2013learning} some Hawkes stuff with learning kernels, test on MemeTracker, lots of math here - READ MORE CLOSELY

\cite{lukasik2015point} log-Gaussian Cox process to infer post frequencies, based on text, look at similar rumours - COME BACK HERE - USING TEXT


\subsection{Other Approaches to Popularity Prediction}

\cite{dou2018predicting} link posts to existing knowledge-base entities, use this as context for popularity prediction; encode KB info in vector, predict with LSTM networks - COME BACK HERE, SIMILAR TO GRAPH INFER?

\cite{he2016deep} predict popularity for news feed selection, natural language and deep reinforcement learning - COME BACK HERE, USING TEXT?

\cite{yang2010predicting} twitter information diffusion based on tweet properties, but mostly user properties - COME BACK HERE, highly cited, needs a closer read

\cite{cheng2014can} predict final size and structure of Facebook image-share cascades using linear SVM model: lots of complicated features, relies on network structure (friendship connections); structural and temporal features most important for prediction - COME BACK

\cite{szabo2008predicting} predict Digg and YouTube final popularity from observation period, three different models: linear regression, constant scaling, growth profile - COME BACK, good comparison/baselines?

\cite{myers2012information} less a prediction and more a diffusion dynamics study - some mention of parameter fit - but mostly proving that there are outside influences, not only diffusion - COME BACK HERE, see if relevant, highly cited

\cite{gruhl2004information} again, not exactly prediction - looking at how information spreads, and the path it follows to do so - highly cited?

\subsection{Node2Vec stuff}

\cite{grover2016node2vec} original node2vec paper, must cite - COME BACK, look at other applications/modifications

\cite{zhou2018efficient} efficiency modifications to existing node2vec implementations - compute transition probs during random walks to reduce memory - as we do

\subsection{Hawkes in general - not exactly relevant?}

\cite{chen2017multivariate} special Hawkes for inhibitory relationships, not mutually-exciting

\cite{delattre2014high} general, fancy Hawkes stuff, very theoretical/mathy

\cite{luo2015multi} again, more general and mathy - multiple triggering patterns

\cite{peng2002multi} software for Hawkes in R

\cite{gao2018transform} meh, probably don't need this - financial application
\cite{gao2017precise} more trading/finance stuff

\cite{halpin2012algorithm} EM Hawkes stuff, and complexity improvements

\cite{roueff2016locally} much math 

\cite{gao2018functional} approximating hawkes processes for a particular application to server queues

\subsection{other maybe-not-related stuff}

\paragraph{mixed weibull}

our model doesn't use a mixed weibull distribution, but I considered it at one time - and logged a bunch of related papers

\cite{razali2013mixture} mixing weibull distributions for better fit of failure times data

\cite{kromer2017accurate} theoretical, fitting mixed weibull distributions

\cite{razali2009combining} mixing parameter for multiple causes of failure

\paragraph{all the rest}

\cite{yuan2018multivariate} network reconstruction of data based on Hawkes processes - no idea how this would tie in at all
}

\section{Conclusions}

Online forums represent a large portion of discussion and communication on the Web. These comment threads allow users to express ideas and opinions that can color the interpretation of the content, making the mechanisms behind the growth of these discussions an important research topic. Existing popularity prediction models, particularly those designed to generate full discussion threads, rely on observations of a post's early comments to predict the remaining activity. This information requirement means these models are incapable of predicting threads for new posts, before comments are present.

In the present work, we introduce our Comment Thread Prediction Model (CTPM), which accurately predicts the size, shape, and temporal dynamics of discussion threads based only on the text of the initial post. This allows for the prediction and simulation of new posts without existing comments. If comments are present, this additional information can be used to improve the model's predictive power. Based on experiments performed on thousands of Reddit discussions, we find that CTPM outperforms existing models, particularly for new posts, or posts with few comments.

\vspace{.2cm}
\noindent\textbf{Future Work.} 
Though the CTPM introduced here consistently outperforms existing models for the prediction of discussion threads, particularly when there are no available comments, there are still avenues for potential improvement. First, CTPM currently uses a simple Jaccard coefficient to quantify the similarity between post titles; a more sophisticated method for determining edge weight in the post graph may potentially improve performance. Another potential area of improvement is in the parameter refinement procedure based on observed comments. The current method takes all observed comments into account, but if the dynamics of the cascade change during its lifetime, the earliest comments may not be a good predictor of the remaining growth. An alternative refinement procedure could give more weight to recent comments, and potentially better recognize the impending death of a cascade. The model could also be tuned specifically to a particular subreddit or domain, perhaps to improve simulation stopping conditions.

The CTPM could also be applied to other social network domains. Can the same mechanisms accurately predict comment threads on Facebook, or retweet cascades on Twitter, without underlying social network knowledge? The potential applications for a generalized popularity prediction model are certainly more widespread than demonstrated in this paper.

Finally, the network embedding techniques employed in the CTPM parameter inference procedure could be applied to other models and problems. The potential benefits of a network-based parameter fit, or embedding values with meaning, should be explored.

\vspace{.2cm}
\noindent\textbf{Acknowledgements.} This research is supported by a grant from the US Defense Advanced Research Projects Agency (DARPA \#W911NF-17-C-0094).

\bibliographystyle{IEEEtran}
\bibliography{bibliography}

\end{document}